\documentclass{aa}  

\usepackage{longtable}
\usepackage{graphicx}
\usepackage{color}

\usepackage{txfonts}
\begin{document} 

   \title{Fan-shaped jet close to a light bridge}

    \author{Y. Liu\inst{1}
            \and
            G.P. Ruan\inst{1}
            \and
            B. Schmieder\inst{2}\fnmsep \inst{3}\fnmsep \inst{4}
            \and
            S. Masson\inst{5}
            \and
            Y. Chen\inst{1}
            \and
            J.T. Su\inst{6}
            \and
            B. Wang\inst{1}
            \and
            X.Y. Bai\inst{6}
            \and
            Y. Su\inst{7}\fnmsep \inst{8}
            \and
            Wenda Cao\inst{9}\fnmsep \inst{10} 
            }

 \institute{Shandong Provincial Key Laboratory of Optical Astronomy and Solar-Terrestrial Environment, and Institute of Space Sciences, Shandong University, Weihai 264209, China\\
            \email{rgp@sdu.edu.cn}
            \and
            Observatoire de Paris, LESIA, Universit\'e PSL, CNRS, Sorbonne Universit\'e, Universit\'e de Paris, 5 place Jules Janssen, F-92190 Meudon, France
            \and
            University of Glasgow,  School of Physics and Astronomy,  Glasgow G128QQ, Scotland, UK
            \and
            Centre for Mathematical Plasma Astrophysics, Dept.of Mathematics, KU Leuven 3001,  Leuven, Belgium
            \and
            Laboratoire de Physique des Plasmas (LPP), Ecole Polytechnique, IP Paris, Sorbonne Universit\'e, CNRS, Observatoire de Paris, Universit\'e PSL, Universit\'e Paris-Saclay, Paris, France
            \and
            Key Laboratory of Solar Activity, National Astronomical Observatories, Chinese Academy of Sciences, Beijing 100012, China
            \and    
            Key Laboratory of Dark Matter and Space Astronomy, Purple Mountain Observatory, Chinese Academy of Sciences (CAS), 8 Yuanhua Road, Nanjing 210034, People$'$s Republic of China
            \and    
            School of Astronomy and Space Science, University of Science and Technology of China, Hefei, Anhui 230026, People’s Republic of China
            \and    
            Center for Solar-Terrestrial Research, New Jersey Institute of Technology, 323 Martin Luther King Blvd., Newark, NJ 07102, USA
            \and    
            Big Bear Solar Observatory, 40386 North Shore Lane, Big Bear City, CA 92316, USA
             }

 
  \abstract
   {}
   {On the Sun, jets in light bridges (LBs) are  frequently observed with  high-resolution instruments. The respective roles played by convection and  the magnetic field in   triggering such jets are   not yet clear.}
  { We report a small   fan-shaped jet  along a LB observed by the  1.6m Goode Solar Telescope (GST)  with the TiO  Broadband Filter Imager (BFI), the Visible Imaging Spectrometer (VIS) in  $H_\alpha$,  and   the Near-InfraRed Imaging Spectropolarimeter (NIRIS), along with the Stokes parameters.  
 The high spatial and temporal resolution of those instruments allowed us to analyze  the features identified during the jet  event. 
   By constructing the H$\alpha$ Dopplergrams, we found  that the plasma is first  moving upward, 
   whereas during the second phase of the jet, the plasma is flowing back.  Working with time slice diagrams, we investigated the propagation-projected  speed   of the fan and its bright base.}
   {The fan-shaped jet developed  within a  few minutes, with diverging beams. At  its base,   a bright point was slipping along the LB and ultimately invaded the umbra of the sunspot. The H$\alpha$ profiles of the bright points enhanced the intensity in the wings, similarly to the case of Ellerman bombs. Co-temporally, the extreme ultraviolet (EUV) brightenings developed at the front of the dark material jet and moved at the same speed as the fan, leading us to propose that  the fan-shaped jet  material compressed and heated the ambient plasma  at its extremities in the corona. }
   {Our multi-wavelength analysis 
   indicates that the fan-shaped jet could result from magnetic reconnection across the  highly diverging field low in the chromosphere,  leading to an apparent slipping motion of the jet material along the LB.
However, we did not find any opposite magnetic  polarity at the jet base, as would typically be expected in such a  configuration. We therefore discuss  other  plausible physical mechanisms, based on waves and convection, that may have triggered the event. }


   \maketitle
%

\section{Introduction}
Solar jets are collimated ejections of plasma in the solar atmosphere. They are a means of mass and energy transport through the heliosphere and, thus, it is important to understand their formation mechanisms. They  commonly consist of  multi-thermal  components of cool and hot plasma, respectively referred to as surges and X-ray jets (see reviews in \citet{Raouafi2016,Shen2021,DePontieu2021,Schmieder2021}).

  The cool plasma of surges ($<  10^{4}$ K)  related to light bridges (LBs)   was first investigated in the 1970s  by \citet{Roy1973}. In a detailed study of four surges combining $H_\alpha$  observations with vector magnetograms, these authors computed the velocity and the length  of the surges and obtained   a maximum  value for the speed  around  $175~\rm{km\,s}^{-1}$ and a maximum length of $50~\rm{Mm}$. After an ascending phase, a descending phase was observed with an acceleration smaller than the gravity. Later, LBs were studied by \citet{Asai2001} using the Transition Region and Coronal Explorer (TRACE), demonstrating smaller ejections ($23~\rm{Mm}$ length)  with a fan-shape as with the jets examined by \citet{Roy1973}.  
  
 The size of such jets spans broad range domains, from $200~\rm{Mm}$  at the edge of active regions to small scale  of $1-2~\rm{Mm}$ that is related to  network and sunspot penumbra. The kinetic energy of jets is related to speed from ${10~\rm{km\,s}^{-1}}$ to ${300~\rm{km\,s}^{-1}}$. Large jets seen in the chromosphere are often  long-lasting and recurring. They have other signatures, such as  brightenings at their  footpoint  in  the chromosphere, as well as hot jets in the transition region and corona. Their associated hot  jets are observed in different wavelengths, from the extreme ultraviolet to the X -ray \citep{Joshi2020}. Over the past decade, thanks to the high spatial and temporal resolution observations  in optical range by advanced telescopes such as the Solar Optical Telescope (SOT, \citet{Tsuneta2008}) on board the Hinode spacecraft, the Interface Region Imaging Spectrograph (IRIS, \citet{DePontieu2014}), the New Vacuum Solar Telescope (NVST, \citet{Liu2014}),  the  Swedish Solar telescope (SST, \citet{Scharmer1991}),  and the  Goode Solar Telescope   (GST- \citet{Goode2012}),  smaller and smaller  jets have been detected  (e.g.,   \citep{Shibata2007,Katsukawa2007,Tian2014,Peter2014,Louis2008,Iijima2017,YangH2019,Ruan2019,Li2020}).

With the development of these  instruments, intense research has been carried out on  active  ejections. 
The chromosphere above sunspots exhibits many various dynamic phenomena such as running penumbra waves, umbra flashes, vortex flows,  and  jets   -- particularly  in cases when the spot has  a fragmented penumbra or a LB 
(e.g; \citet{Louis2008,YangH2019,Asai2001,Robustini2016,Robustini2018,Li2020,Lim2020}).  It is known that LBs have a weaker magnetic field compared to the surrounding umbra, with the magnetic field lines  more inclined over the solar surface  than in the umbra \citep{Toriumi2015a,Toriumi2015b}.  With the SOT on board of Hinode, small jets were observed by the wide-band  pass Ca II filter  (e.g., \citep{Shimizu2009,Louis2014}).  \citet{Shimizu2009} reported  on a sunspot LB that produced chromospheric plasma ejections intermittently and recurrently for more than one day. It has been explained that EUV jets and $H_\alpha$ surges represent hot and cool plasma ejections, respectively, along different field lines because they are not co-spatial at high spatial resolution \citep{Chae1999}.  The jets above LB may  reach up to the lower corona and the front  is heated up to transition region and  coronal temperatures \citep{Bharti2015,Yuan2016}. Long  spicules  (sometimes referred to as  small jets) are  observed in $H_\alpha$  and may provide hot plasma to the corona and most of the enhanced spicules are thought to channel hot plasma into the corona \citep{Samanta2019}. 
 
The magnetic configuration of LBs shows frequently opposite magnetic polarities compared to the surrounding umbra  \citep{Toriumi2015a,Bharti2007,Lim2020}. Therefore   the most   well  known model of jet formation is energy release by magnetic reconnection in the chromosphere \citep{Heyvaerts1977,Shibata2007,Chae1999,Shimizu2009}. \citet{Tian2018,Lim2020} found opposite magnetic polarities at the light bridge, opening the possibility of a reconnection between the emerging flux and the surrounding sunspot field. But  the  trigger of the acceleration is not well understood.  The cool material can be accelerated by magnetic tension of reconnected field lines and create a slingshot effect \citep{Nishizuka2008}, or by slow shock waves colliding with the transition region \citep{Shibata1982,Takasao2013}.   In fact, \citet{Shibata1982} distinguished two types of macrospicules depending on the density  of the shock regions: shocks in the transition region or shock waves originate in network bright points in the photosphere, with the latter potentially creating shock-tube type jets.  

Opposite  polarities are  sometimes difficult to   evidence in LBs.  \citet{Bai2019} studied a fan-shaped jet caused by  convective motions in   the photosphere and could detect parasitic polarity at the jet base after applying a special treatment  to the Stokes parameter profiles  as well as \citet{Franz2013}. In these cases, the authors found  abnormal Stokes V profiles revealing a configuration of three lobes. The  detection of  a third lobe  in the Stokes V profiles demonstrates the presence of opposite polarity.  The detection of three-lobe profiles in their  analysis induced an increase of  the area of opposite polarity   in the penumbra from ${4\%}$ to ${17\%}$. 

Another possible way of triggering jets is from the MHD waves from the photosphere, whereby shocks could be  formed from magneto-acoustic waves caused by the leakage of p-mode oscillations   or  by Alf\'ven waves that  push the plasma inside the flux tubes by increasing the  pressure in magnetic concentration  like in macrospicules \citep{Shibata1982}.  In recent simulations,  chromospheric jets  were   produced by shock waves passing through the transition region. \citet{Iijima2015} found that the scale  of chromospheric jets is related to the temperature of the  ambient corona. By using two-dimensional (2D) magnetohydrodynamic simulations, they found that  shorter jets exist in relatively cooler corona, as in coronal holes.  Various  excited MHD waves could  produce chromospheric jets in three-dimensional (3D) simulations  \citep{Iijima2017}. The results show that the simulated jets are closely related to  solar spicules. Later on, it was shown that  numerical simulations concerning  the Sun compared to  laboratory fluid dynamics experiments could explain the formation mechanism of the jets. Under  the effects of gravity, the non-linear focusing of quasi-periodic waves in an  anisotropic media of both magnetized plasma and polymer fluid is sufficient to generate a large number of jets  similar to chromospheric spicules \citep{Dey2022}. This MHD wave domain should be investigated  further in future to explore the initiation of jets.  

At the base of surges, brightenings  in chromospheric lines have been observed. Studies have found that the footpoint brightening of plasma ejection from a LB  is  similar to  Ellerman bombs \citep{Bharti2007}. Their  line profiles have been compared with Ellerman bombs (EBs) or IRIS bombs  (IBs)  
\citep{Yang2019,Joshi2021}. An EB has enhanced emission in  the $H_\alpha$ wings  and  commonly no signature in $H_\alpha$  line core  \citep{Ellerman1917,Rezaei2015,Grubecka2016,Chen2019}.  It is thought to be  triggered by magnetic reconnection in the lower atmosphere  from 50 to $900~\rm{km}$ above the solar surface \citep{Grubecka2016}.  On the other hand,  brigthenings in the umbra, called umbra dots have been explained by convection bringing  above the solar surface hotter plasma and could be responsible for the LB formation \citep{Katsukawa2007}.  
LBs in a sunspot favor the  decay of the spot and its  submergence by convective flows. 
This idea has been also investigated in the sunspot formation simulation of \citet{Rempel2012}, where is was shown that hot bubbles from the convection may invade umbra and penumbra, serving as tools for fragmenting  sunspots.

There are several major debates centered around the formation of tiny jets in sunspot LB,  particularly regarding the role played by   magneto-convection and by magnetic reconnection.

In this paper, we present high-resolution observations  of a fan-shaped jet obtained by the Goode Solar Telescope (GST)  operating at  the Big Bear Solar Observatory (BBSO),  as well as  by  the  Atmospheric Imaging Assembly (AIA) aboard  the Solar Dynamic Observatory (SDO). We consider the dynamic evolution of the jet flow  and the vector magnetograms  in Section 2. In Section 3, we summarize our results  and  discuss  the implications of our observations on   explaining the trigger mechanisms of LB jets. 

\begin{figure*}[!htbp]
\begin{minipage}{\textwidth}
\centering
\includegraphics[width=50mm,angle=0,clip]{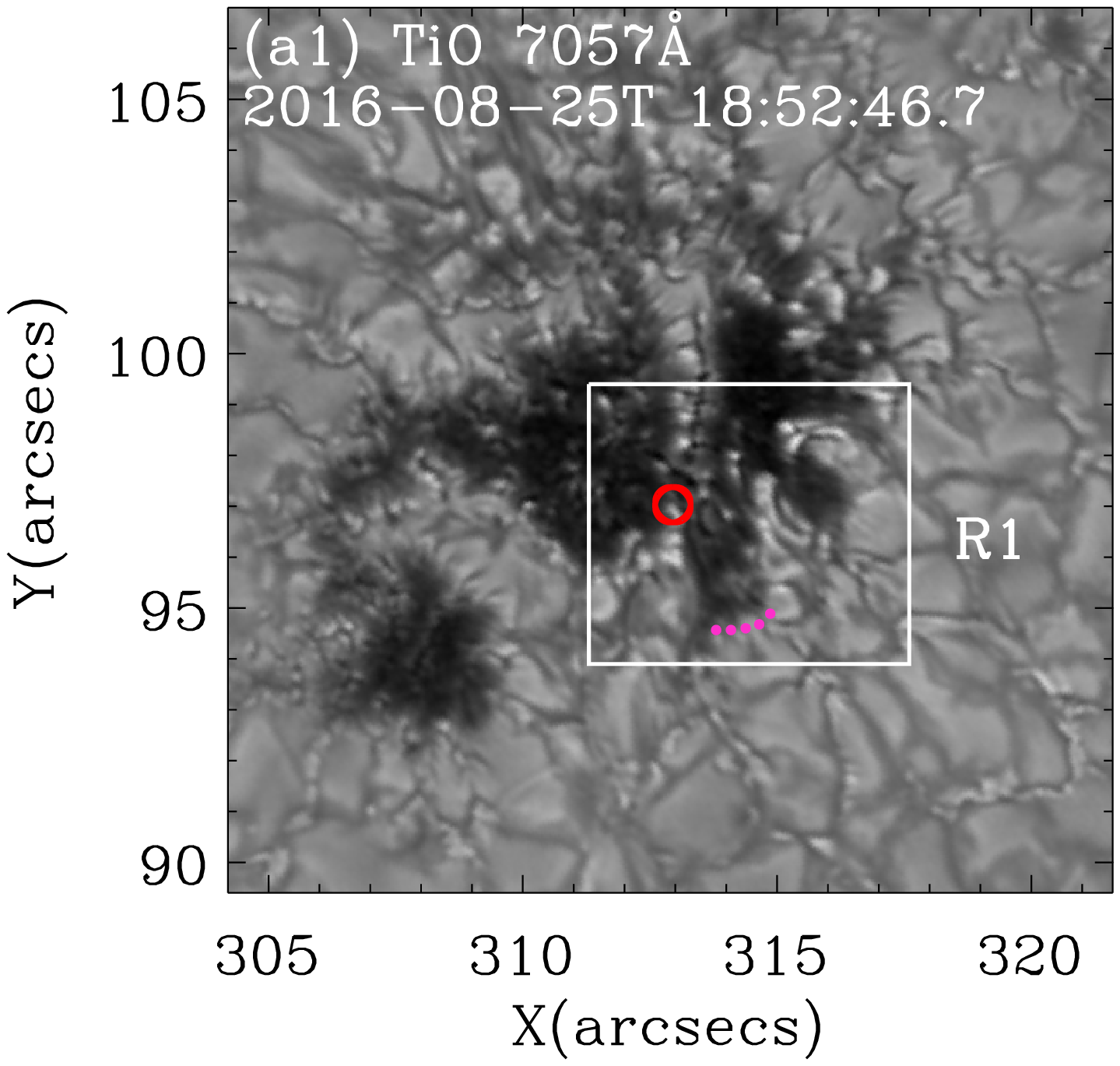}
\includegraphics[width=50mm,angle=0,clip]{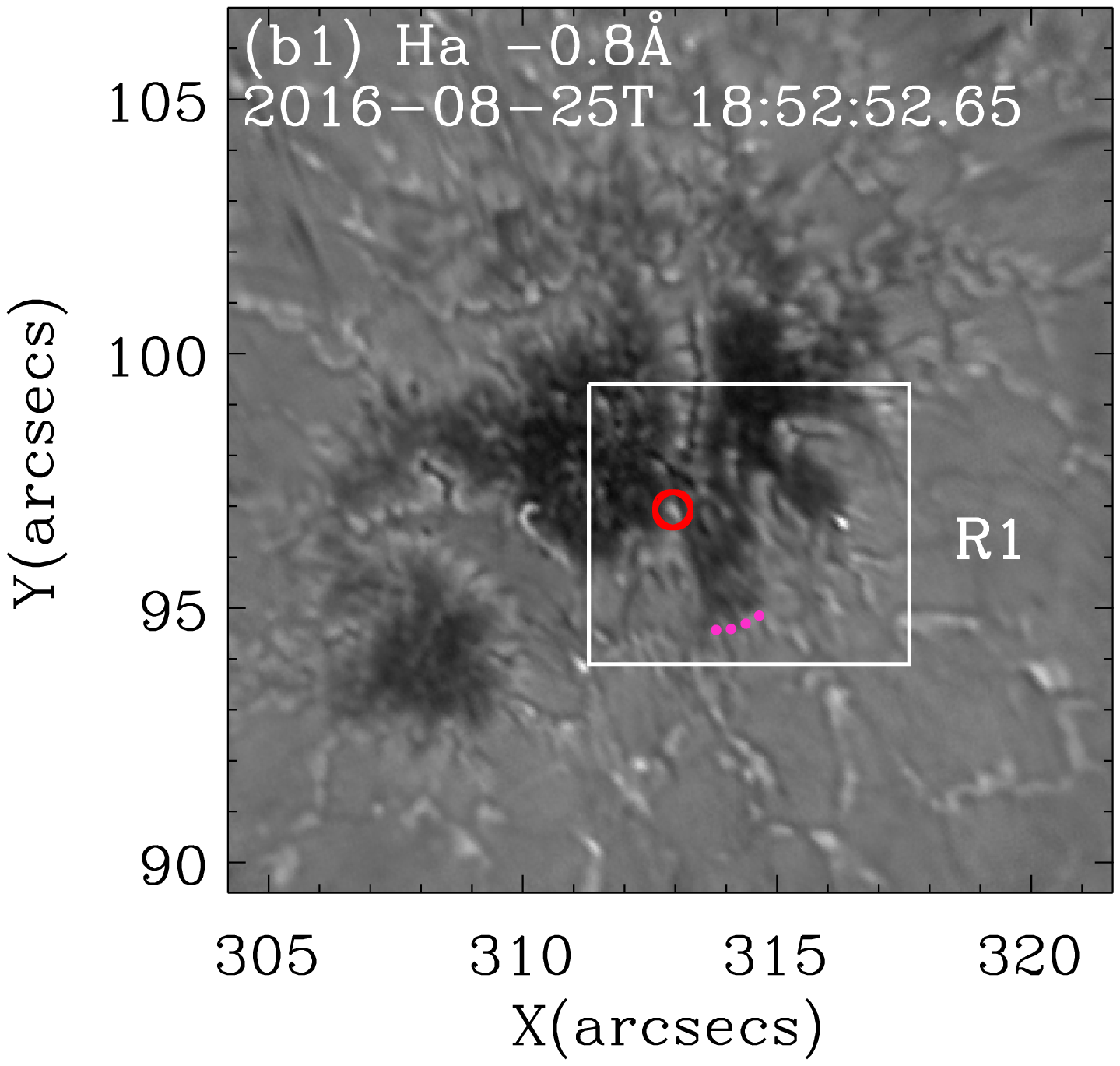}
\includegraphics[width=50mm,angle=0,clip]{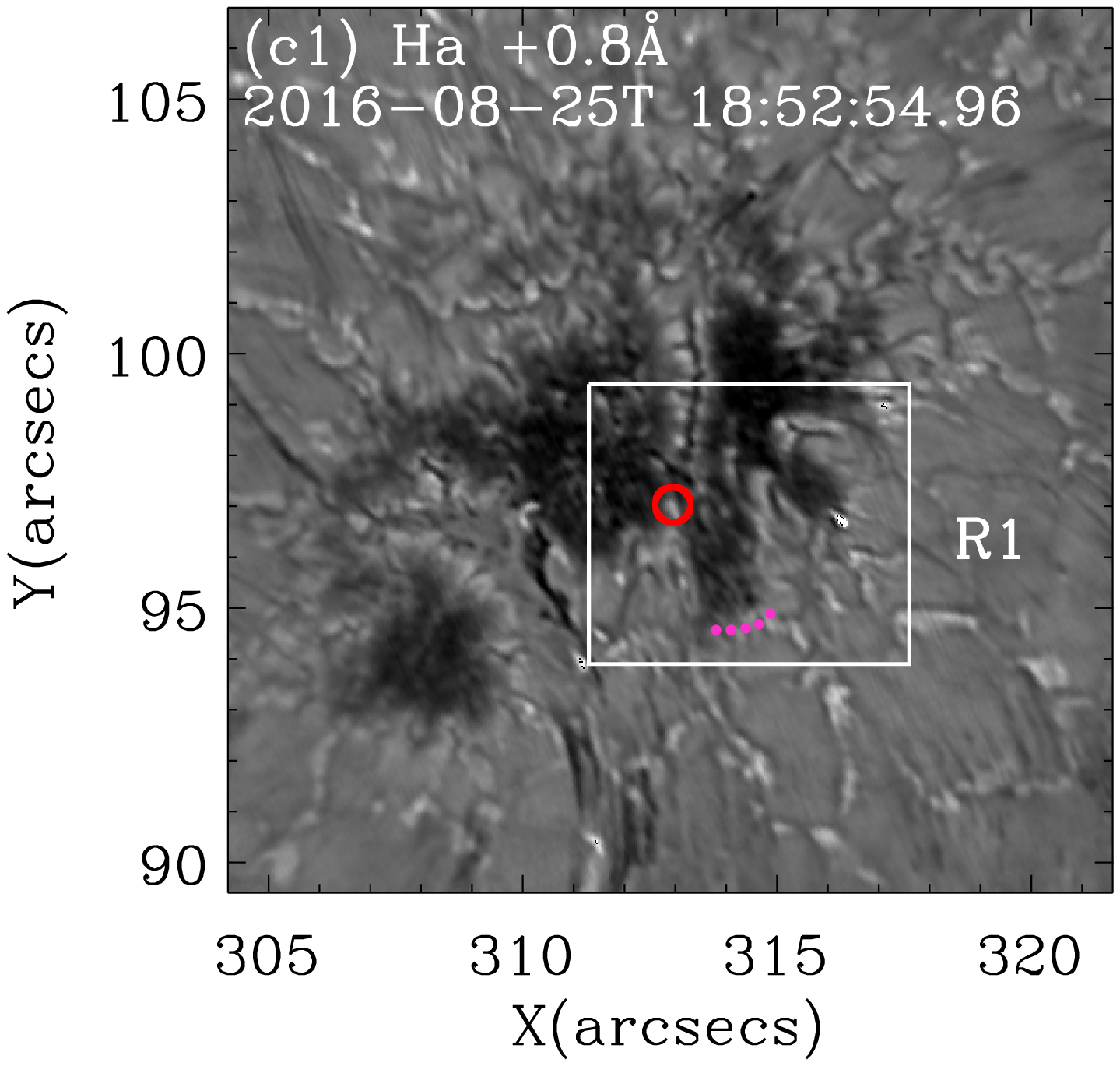}
\includegraphics[width=50mm,angle=0,clip]{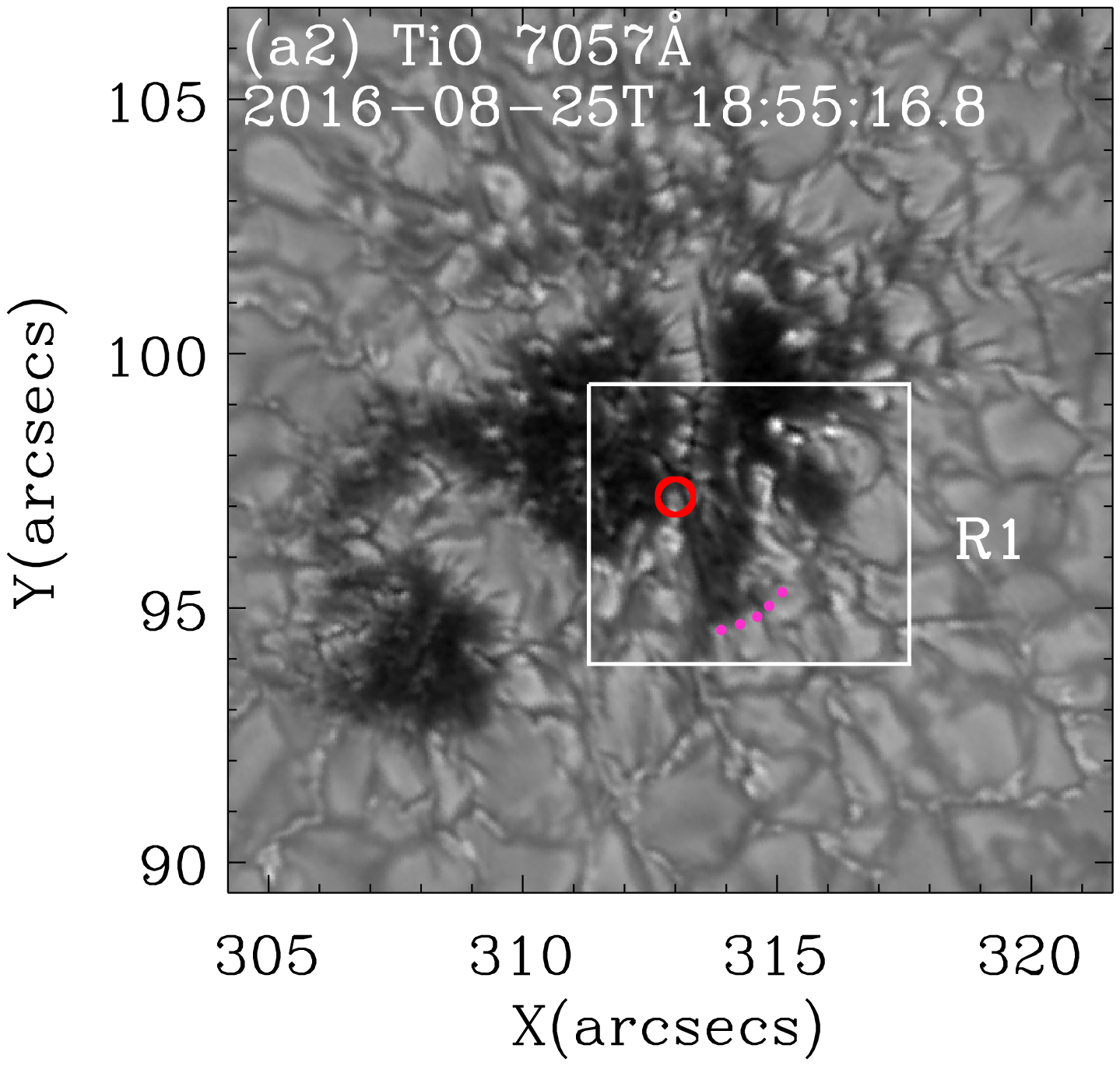}
\includegraphics[width=50mm,angle=0,clip]{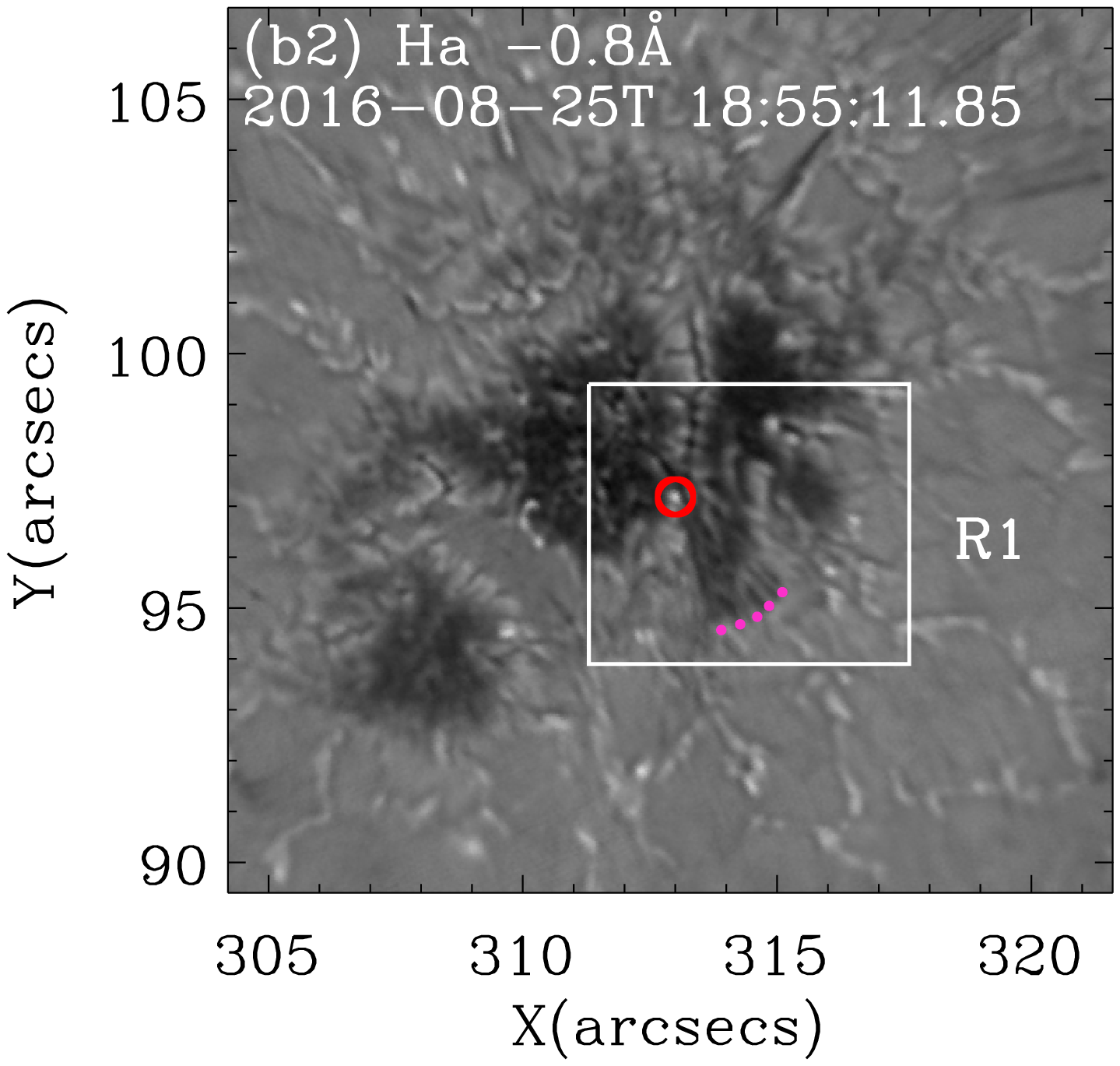}
\includegraphics[width=50mm,angle=0,clip]{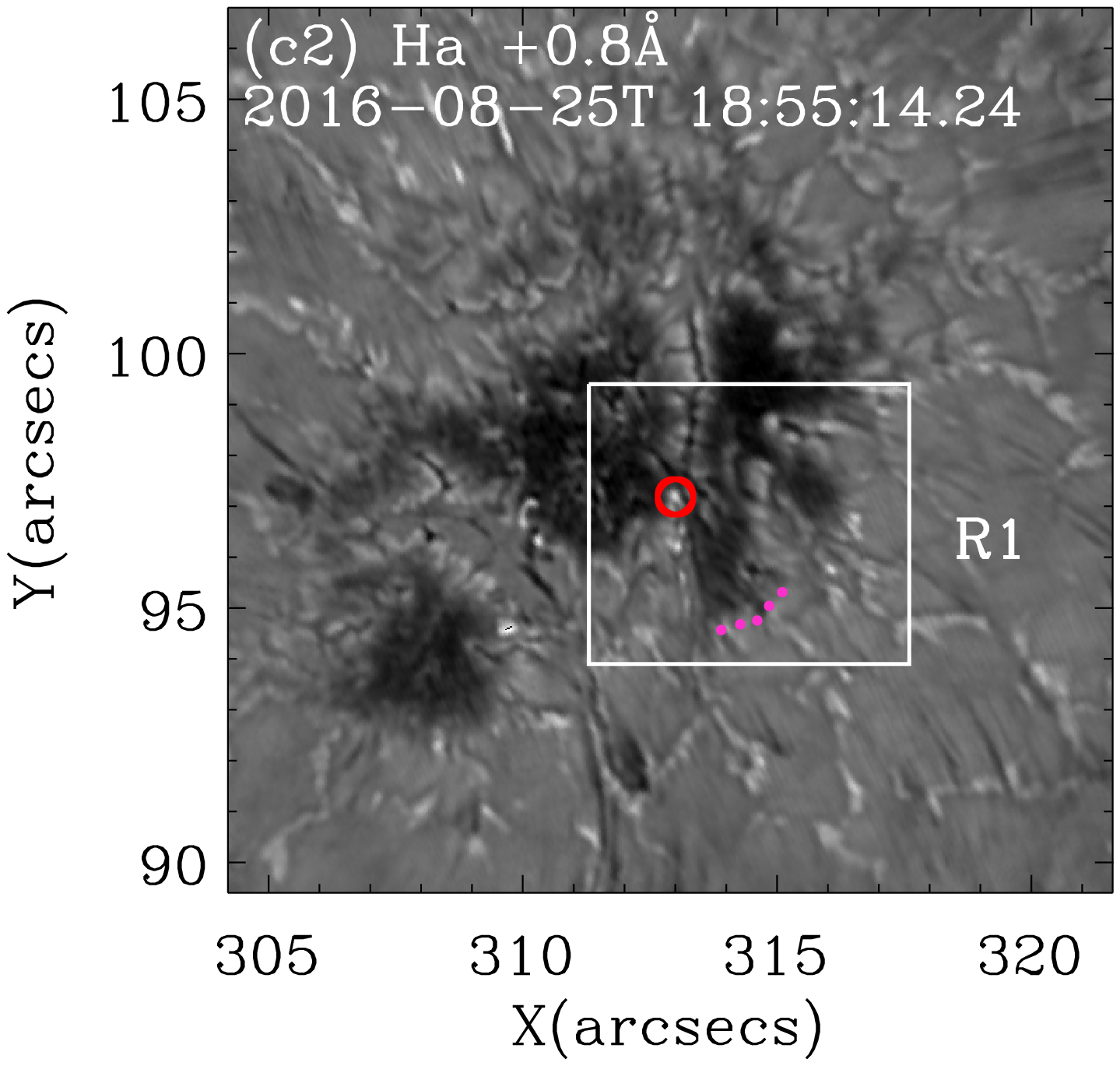}
\includegraphics[width=50mm,angle=0,clip]{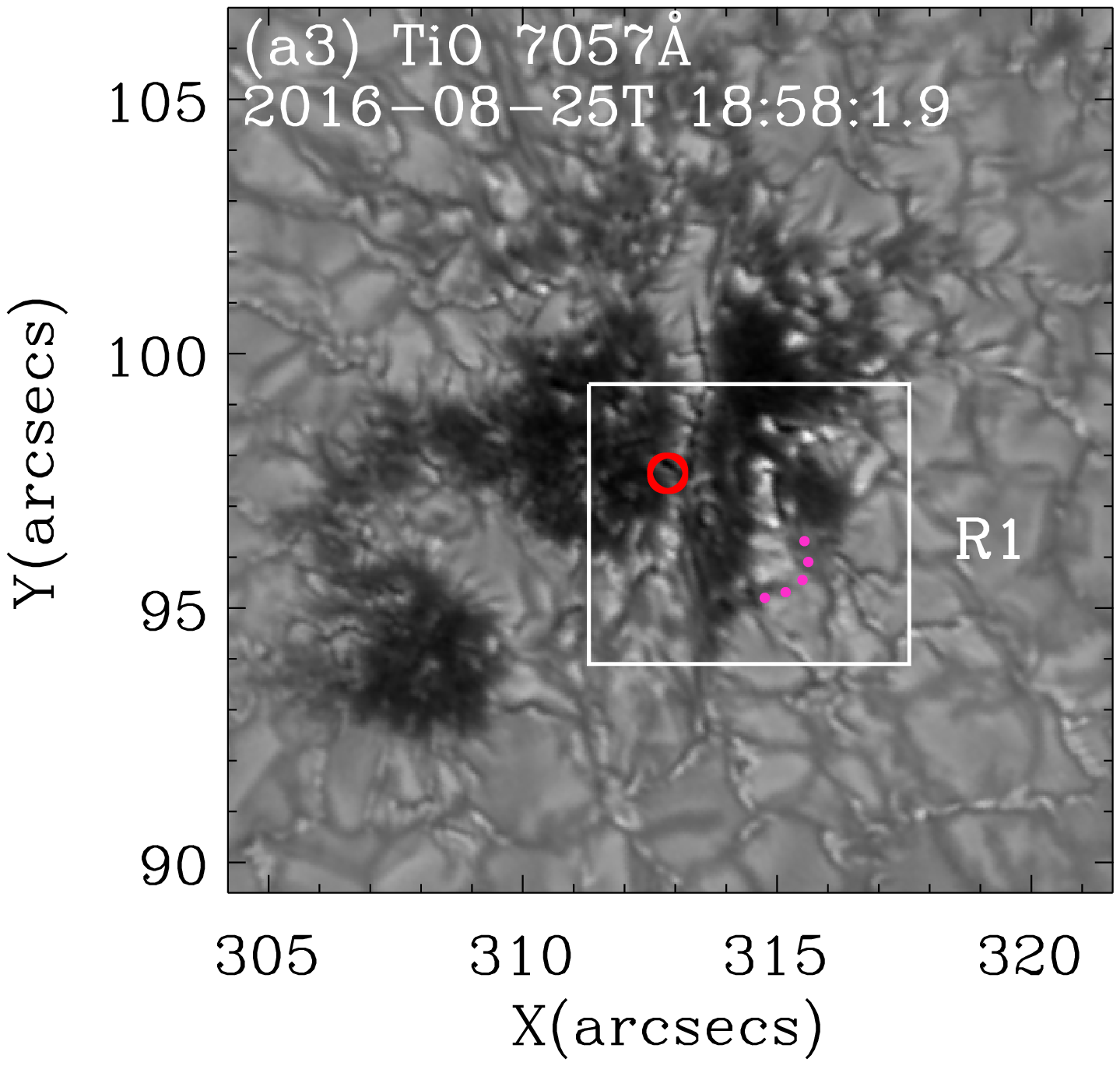}
\includegraphics[width=50mm,angle=0,clip]{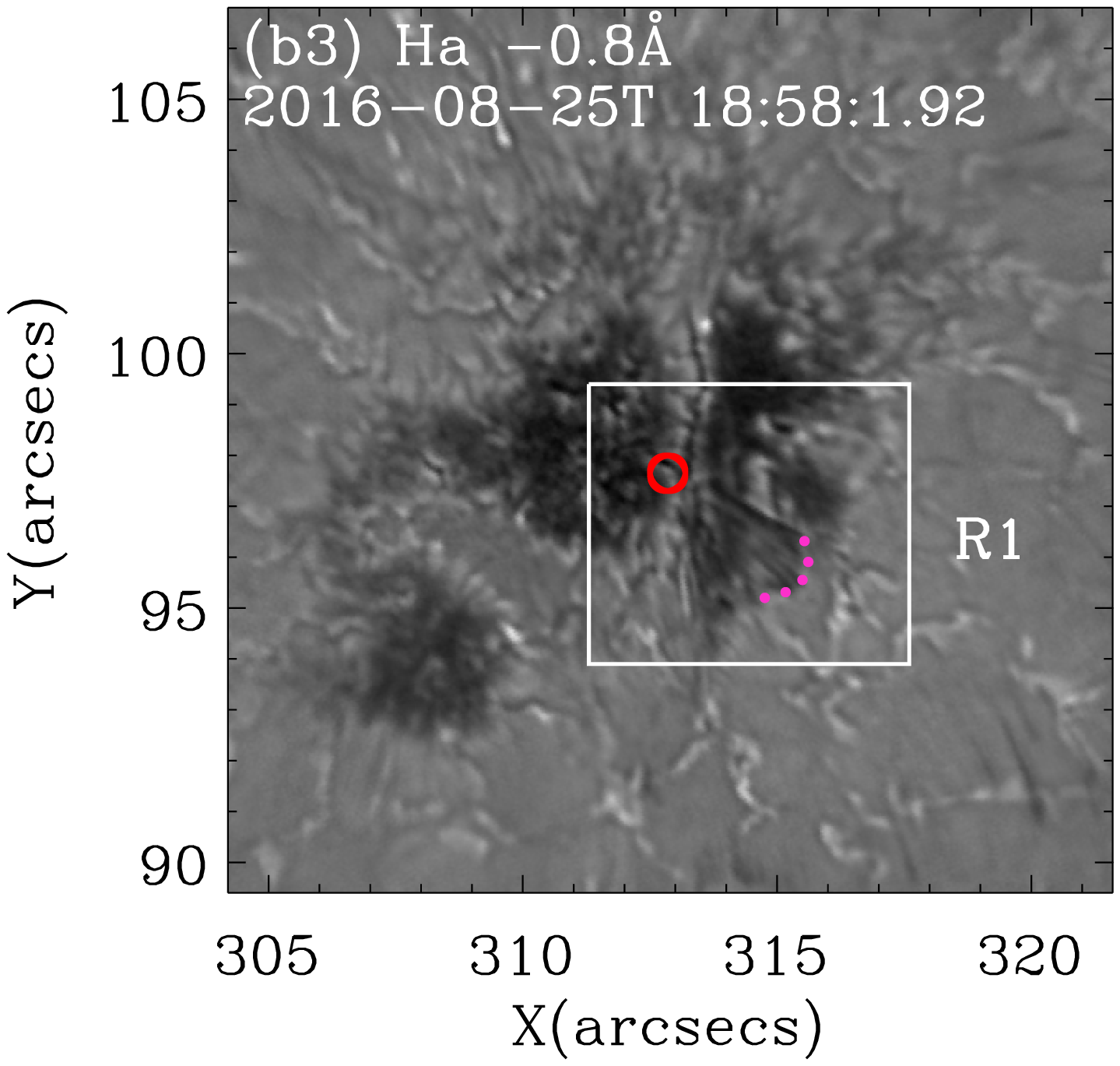}
\includegraphics[width=50mm,angle=0,clip]{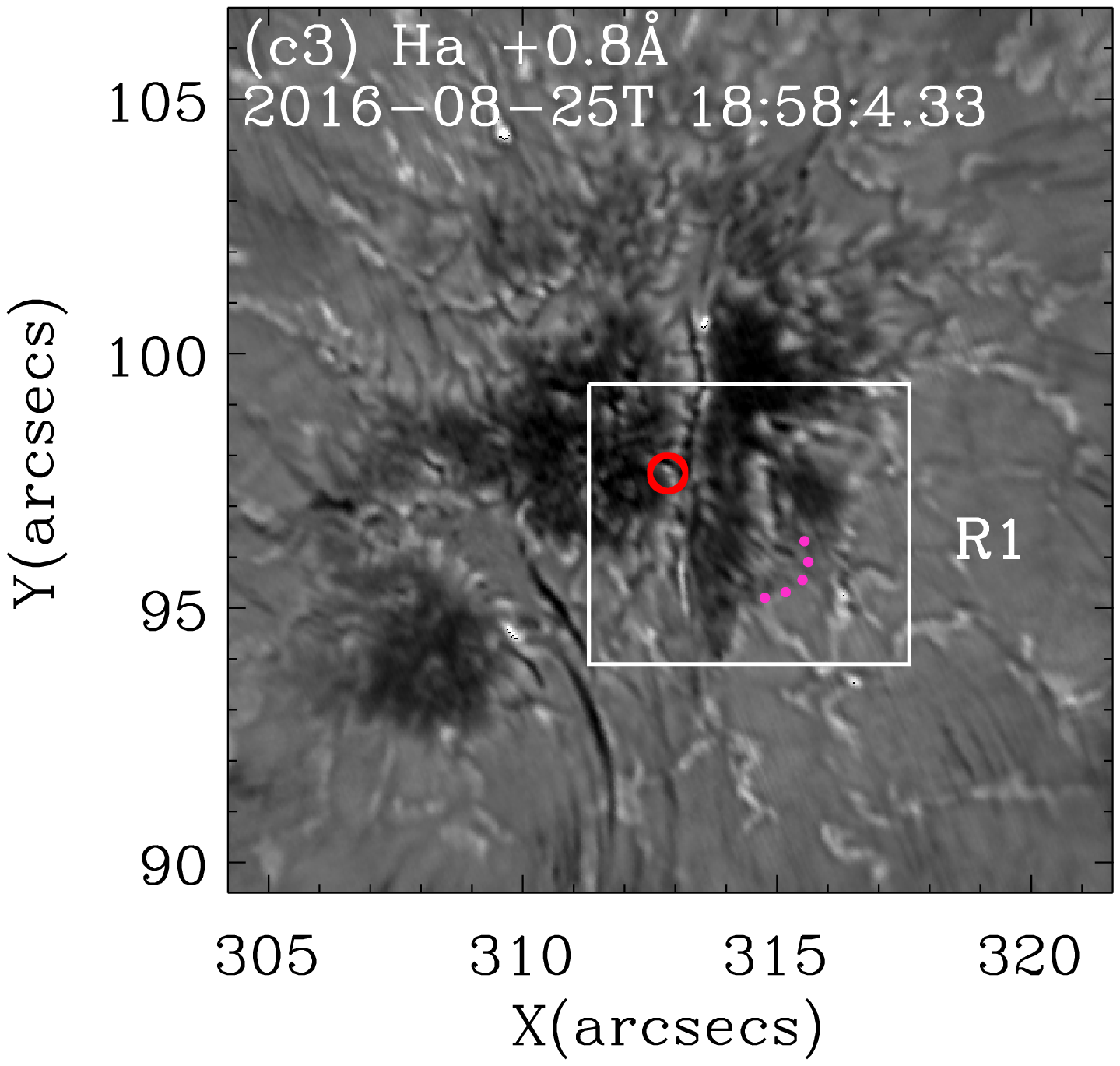}
\includegraphics[width=50mm,angle=0,clip]{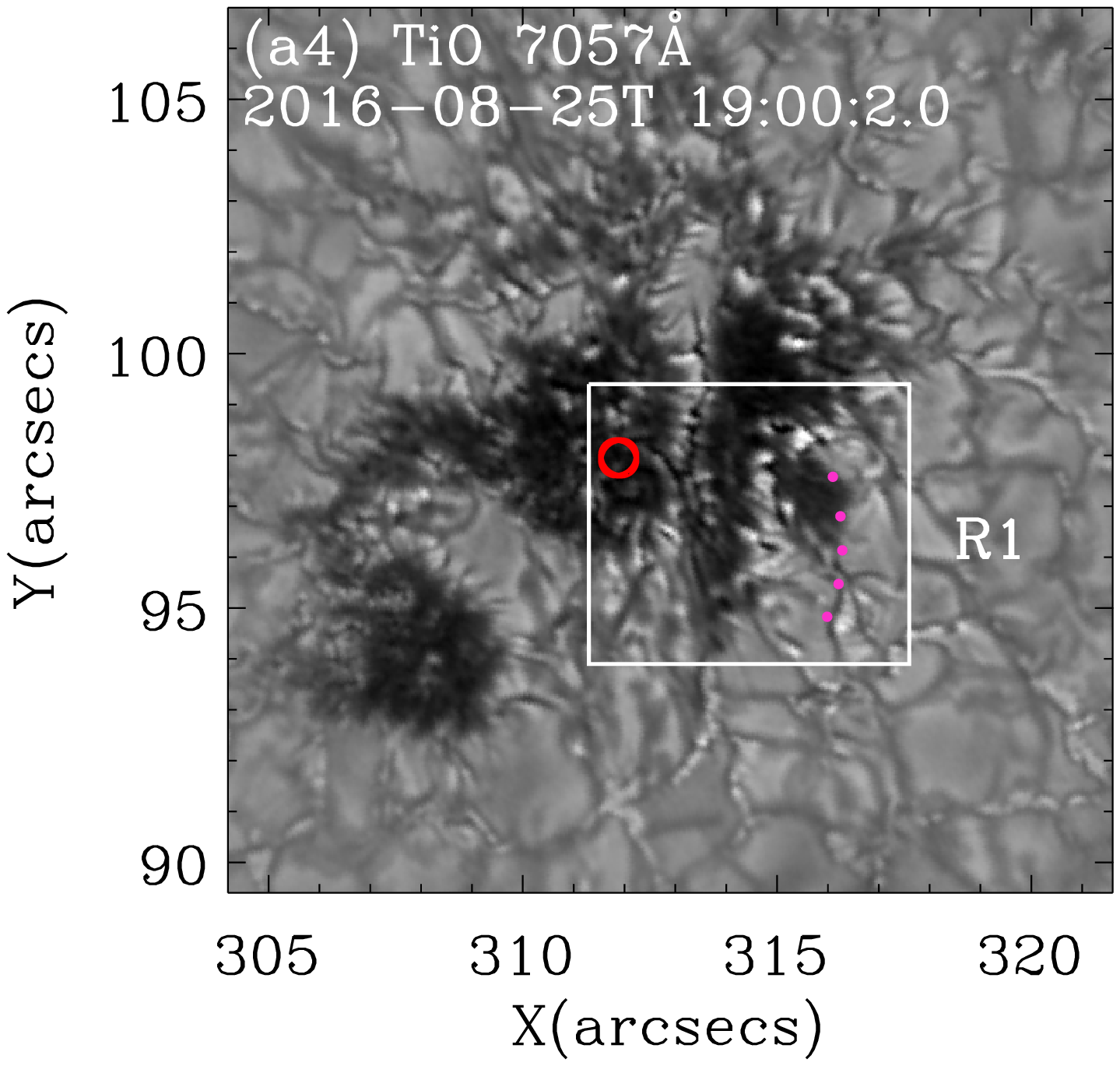}
\includegraphics[width=50mm,angle=0,clip]{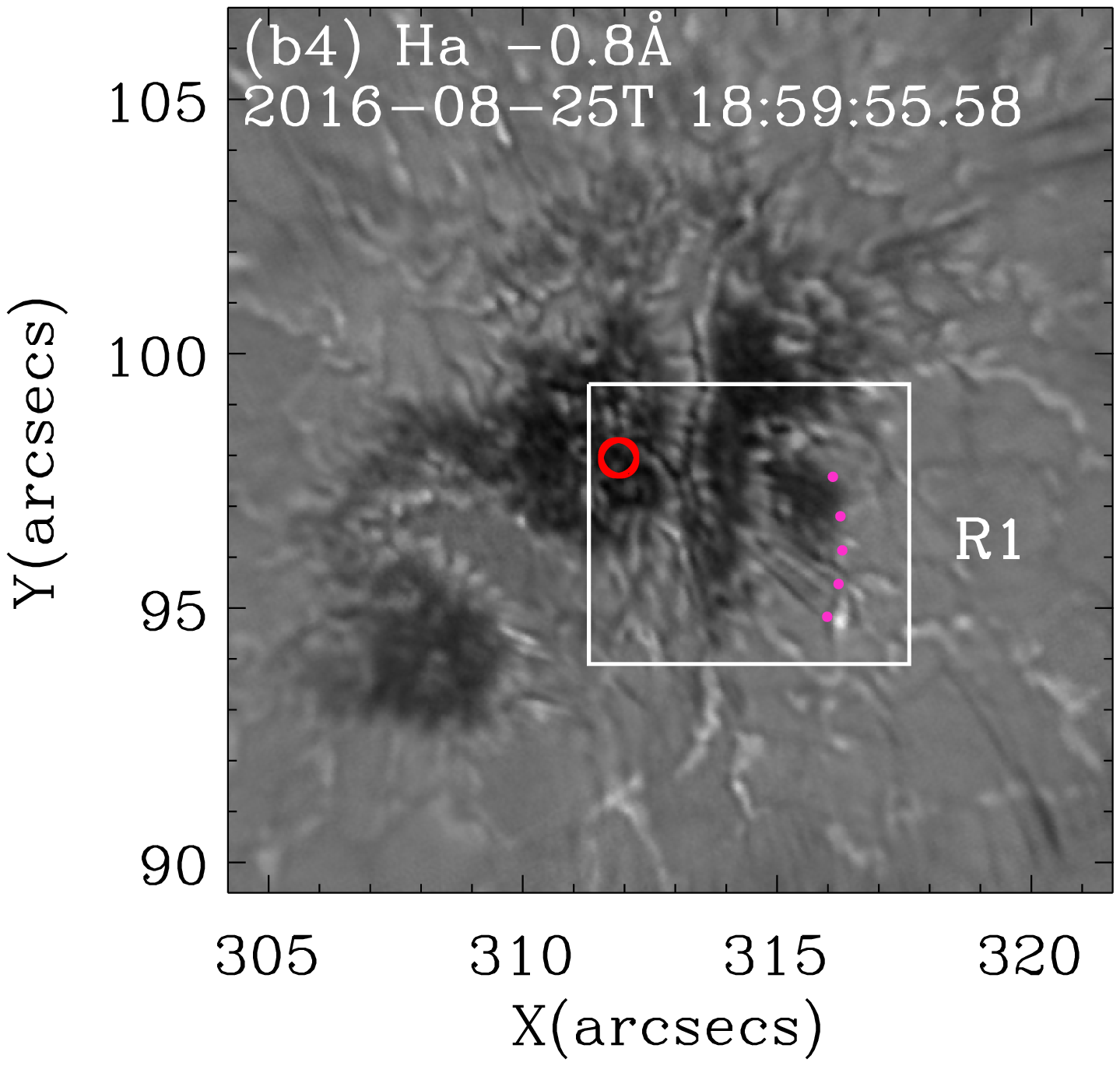}
\includegraphics[width=50mm,angle=0,clip]{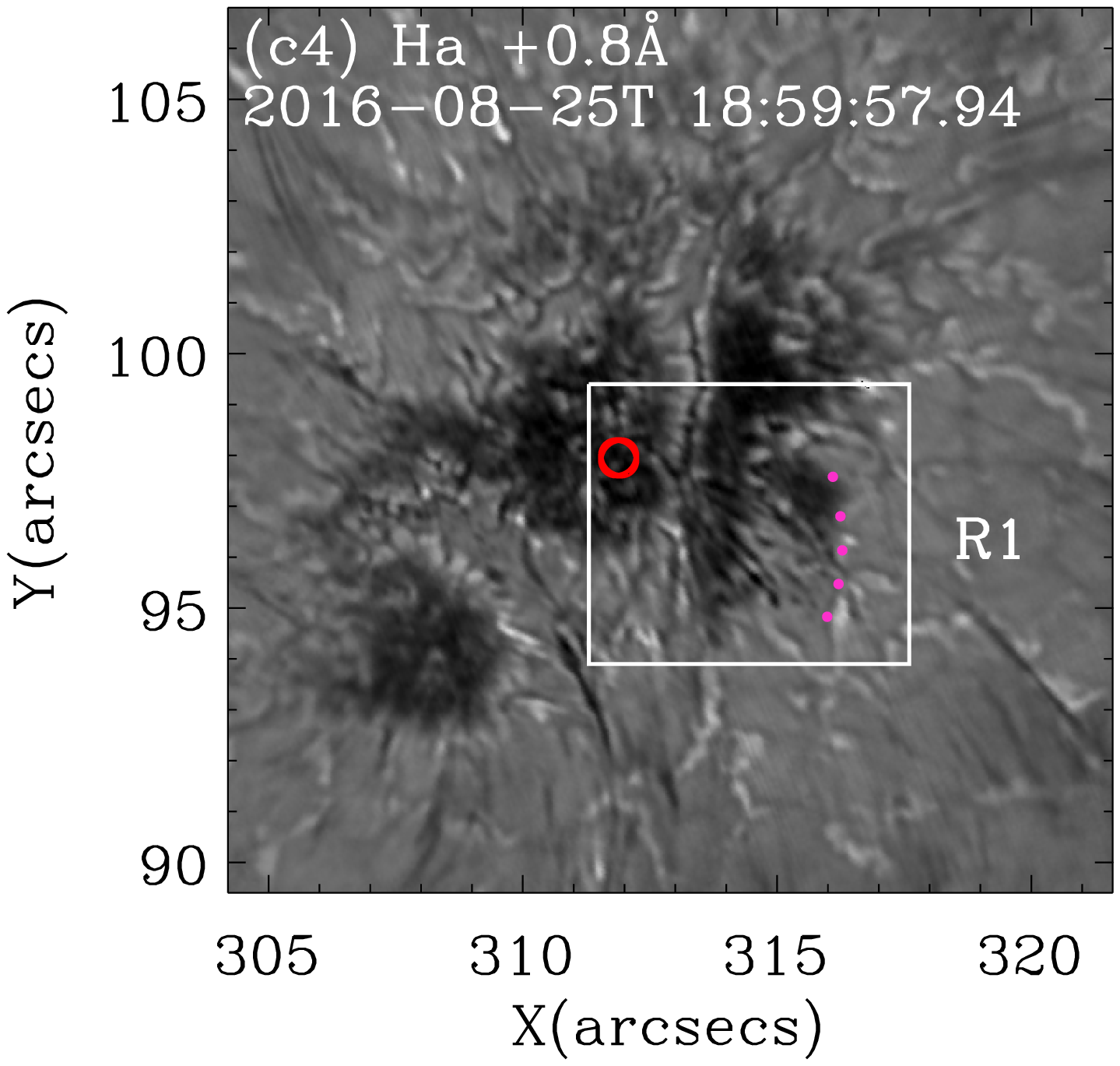}
\end{minipage}

\caption{ Temporal  evolution of a fan-shaped jet observed in the active region NOAA  12579   with the GST at BBSO  on  August 25 2016 between  18:52 UT and 18:59 UT. The different panels show: the TiO images (panels a1, a2, a3, a4), the $H_\alpha$ blue wing   images at -0.8 \AA{} (panels b1, b2, b3, b4), and $H_\alpha$ red wing images at +0.8 \AA{}   (panels c1,c2, c3, c4).  All H$\alpha$ images are  obtained with the  GST/VIS. R1 represents the area where the jet occurred (FOV of Figures \ref{slice_diagram} and 4). The footpoint and front of the jet are marked by a red circle and magenta dots, respectively.  
}
\label{images_BBSO}
\end{figure*}

\section{{Observations}}
\label{observation}
\subsection{Instruments}
On August 25, 2016, a fan-shaped jet 
 was observed in the leading sunspot of the NOAA active region AR 12579 located at  N11W22 \footnote[1]{https://www.solarmonitor.org/} with the Big Bear Solar Observatory (BBSO) coupled with the 1.6 meter Goode Solar Telescope (GST) \citep{Goode2012} as well as the Solar Dynamic Observatory (SDO) \citep{Pesnell2012} coupled with the Atmospheric Imaging Assembly (AIA) \citep{Lemen2012}. The pointer of the GST  was centered on  the leading spot of the active region. We focused  the field of view on  the LB in  a small  field of view  around 17 $^{\prime \prime} \times 17 ^{\prime \prime}$.

The GST data contain simultaneous observations of the photosphere, using the titanium oxide (TiO) line taken with the Broadband Filter Imager, and the chromosphere, using the $H_\alpha$ 6563~ \AA{} line obtained with the Visible Imaging Spectrometer (VIS) \citep{Cao2010}. The passband of the TiO filter is 10~\AA{}, centred at 705.7~nm, while its temporal resolution is about 15 s with a pixel scale of 0.$^{\prime\prime}034$. Concerning  the VIS, a combination of 5~\AA{} interference filter and a Fabry-P{\'e}rot etalon is used to get a bandpass of 0.07~\AA{} in the $H_\alpha$ line. The VIS  field of view (FOV) is about 70$^{\prime\prime}$ with a pixel scale of $0.^{\prime\prime}029$. To obtain more spectral information, we scan the $H_\alpha$ line at 11 positions with a 0.2\AA{} step following this sequence: $\pm$ 1.0, $\pm$ 0.8, $\pm$ 0.6, $\pm$ 0.4, $\pm$ 0.2, 0.0 \AA{}. We obtained a full Stokes spectroscopic polarimetry using the Fe I 1565 nm doublet over a 85$^{\prime\prime}$ round FOV with the aid of a dual Fabry-P{\'e}rot etalon by the NIRIS Spectropolarimeter \citep{NIRIS;Cao2012}. Stokes I, Q, U, and V profiles were obtained every 72 s with a pixel scale of 0.$^{\prime\prime}$081. All TiO and $H_\alpha$ data were speckle reconstructed using the Kiepenheuer-Institute Speckle Interferometry Package \citep{Woger2008}.

We analyze the EUV multi-wavelength imaging data from SDO/AIA, the vector magnetic field, and continuum intensity data by the Helioseismic and Magnetic Imager (HMI) \citep{Schou2012} on board the SDO spacecraft. AIA observes the Sun in ten different wavebands, covering a wide range of temperatures and reveals physical processes at various layers of the solar atmosphere. The pixel size of the data is $0.^{\prime\prime}6$ and the temporal cadence is  $12~\rm{s}$. Generally, HMI provides four main types of data: dopplergrams (maps of solar surface velocity), continuum filtergrams (broad-wavelength photographs of the solar photosphere), and both line-of-sight and vector magnetograms (maps of the photospheric magnetic field). The processed HMI continuum intensities and magnetograms data are obtained with a  45~\rm{s} cadence and a $ 0.^{\prime\prime}6 $  pixel size, provided by the HMI team. Here, we mainly used   the data from seven EUV channels of AIA and HMI continuum intensity.  For comparison with NIRIS we analyzed the HMI magnetograms in the 24 hours before the event. Continuum intensity maps of HMI  help us to co-align the  TiO, $H_\alpha$ images  and the magnetograms taken by GST. The GST images taken at each wavelength position were internally aligned using the cross-correlation technique  provided by the BBSO programmers. The co-alignment between SDO/HMI  continuum and GST images was achieved by comparing commonly observed features of sunspots in Fe I 6173 \AA{} images and TiO, $H_\alpha$ +/-  0.8  \AA{} images taken frame by frame.

 \subsection{Fan-shaped jet}
 \label{fan-shaped jet}
 
 In the present work, we are interested in a fan-shaped jet invading the umbra of the leading negative sunspot of AR 12579  observed in the $H_\alpha$ lines with the GST. The event occurs between 18:42:43 UT and  19:16:00 UT.
 Figure 1 shows the temporal evolution of the event for three  wavelengths: TiO, $H_\alpha$ -0.8 \AA{} and  $H_\alpha$ + 0.8 \AA{} at four different times.

 The dark structure of the fan-shaped jet that was initially identified was observed at 18:52:46 UT. Its  footpoint is located in the sunspot at  $[x,y] = [313^{\prime\prime},97^{\prime\prime}]$  and the fan$'$s extremities are located between  $[x,y] = [313.8^{\prime\prime},94.6^{\prime\prime}]$ and $[x,y] = [314.8^{\prime\prime},94.9^{\prime\prime}]$ in the $H_\alpha$ blue wing at -0.8 \AA{} (Figure 1, panel b1). During this first phase, between  18:52:46 UT and 19:00:23 UT, the fan$'$s beams display a sweeping motion from the south toward the north (Figure 1, left column).  The extension of each beam of the jet is around  2.5 arcsec, close to 2000 km, and the width of each beam is about 21 km.  Prior to this jet appearing, there were other  mini jets but lasting only  for 4-5 minutes, so we do not analyze them here. 

We computed the height-time diagram   of the fan-shaped jet northward edge to determine its  projected propagation speed. It first propagates at $3.15~\rm{km\,s}^{-1}$ until $\simeq$ 18:57 UT and then it accelerates to reach  $6.91~\rm{km\,s}^{-1}$ until 19:00 UT (Figure 2, top-right  panel). At 18:59:55 UT the fan extends from  $[x,y] = [316^{\prime\prime},94.8^{\prime\prime}]$ to $[x,y] = [316.1^{\prime\prime},97.6^{\prime\prime}]$. At this time, we can also identify the fan-shaped jet in the $H_\alpha$ red wing at +0.8 \AA{}.

 A few seconds later, at 19:00:23 UT, the fan-shaped jet disappears in the $H_\alpha$ blue wing at -0.8 \AA{} but it remains visible in the $H_\alpha$ red wing at +0.8 \AA{}.  Between 19:00:23 UT and 19:03:45 UT the dark material forming the fan-shaped jet seems to flow back down into the sunspot toward the footpoint. While in the first phase, the size of the fan increases with time, in the second phase, the fan$'$s extension decreases with time.

\begin{figure*}[!htbp]
\begin{minipage}{0.5\textwidth}
\centering
\includegraphics[width=80mm,angle=0,clip]{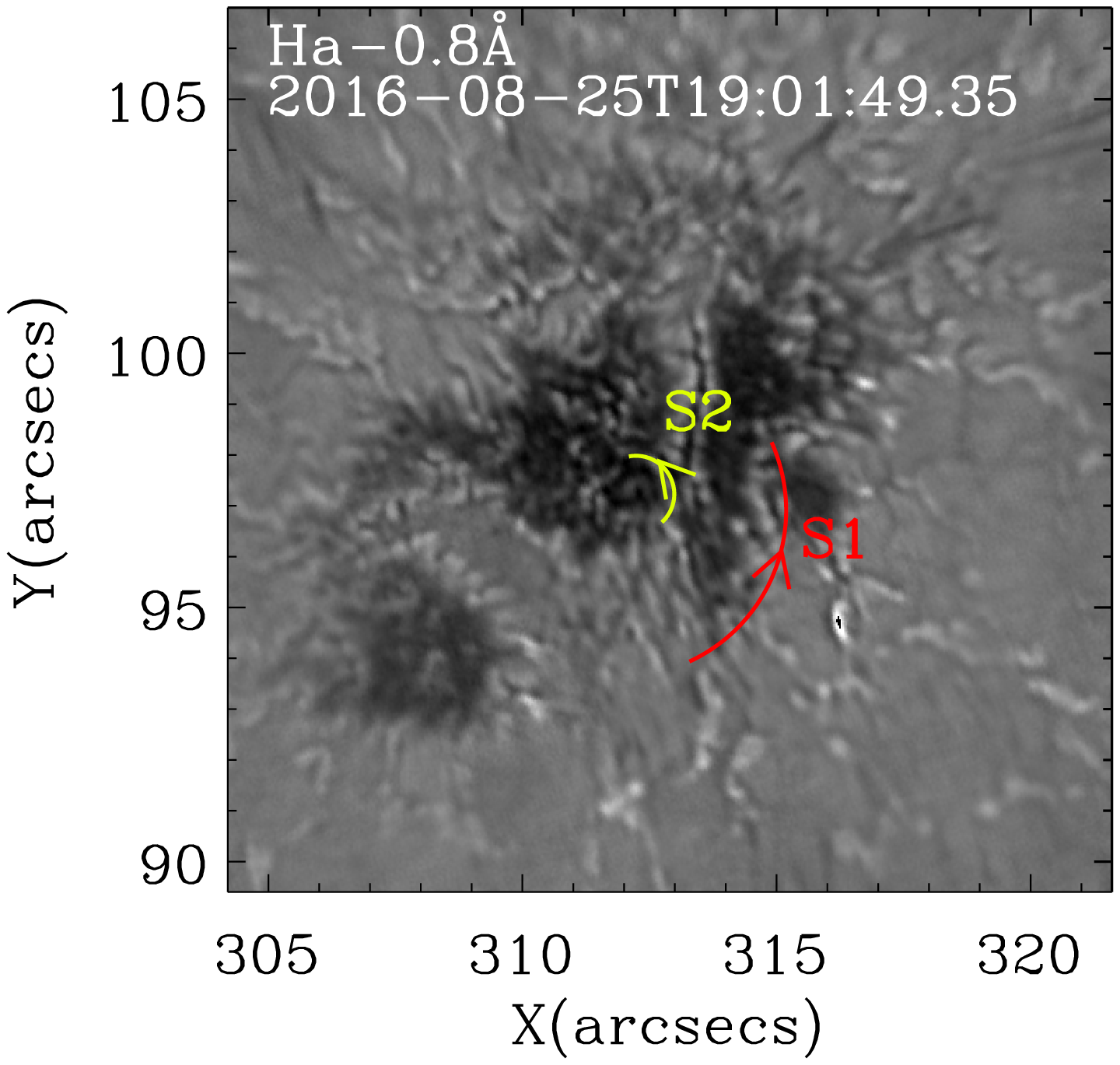}
\end{minipage}
\begin{minipage}{0.5\textwidth}
\centering
\includegraphics[width=90mm,angle=0,clip]{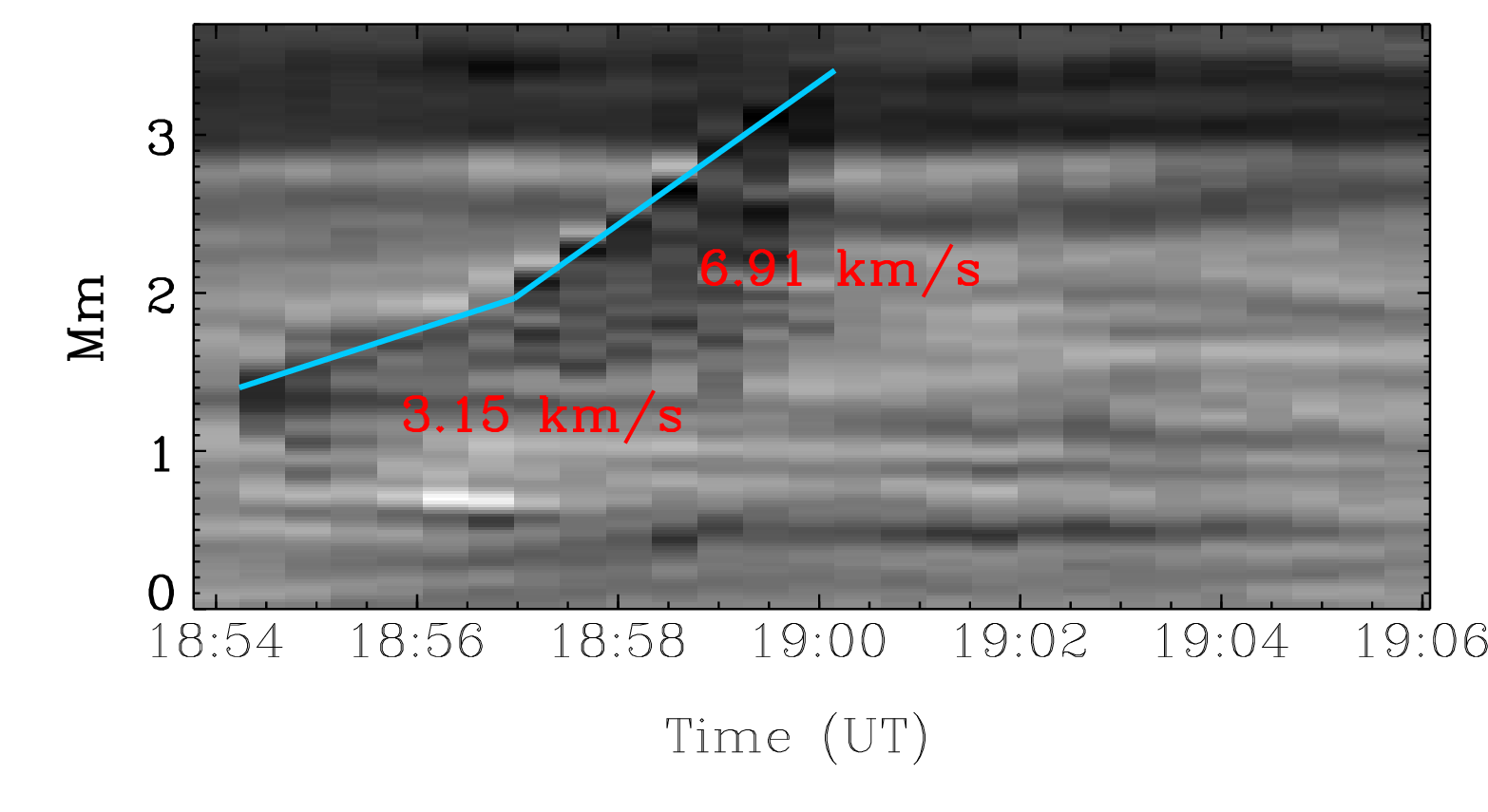}
\includegraphics[width=90mm,angle=0,clip]{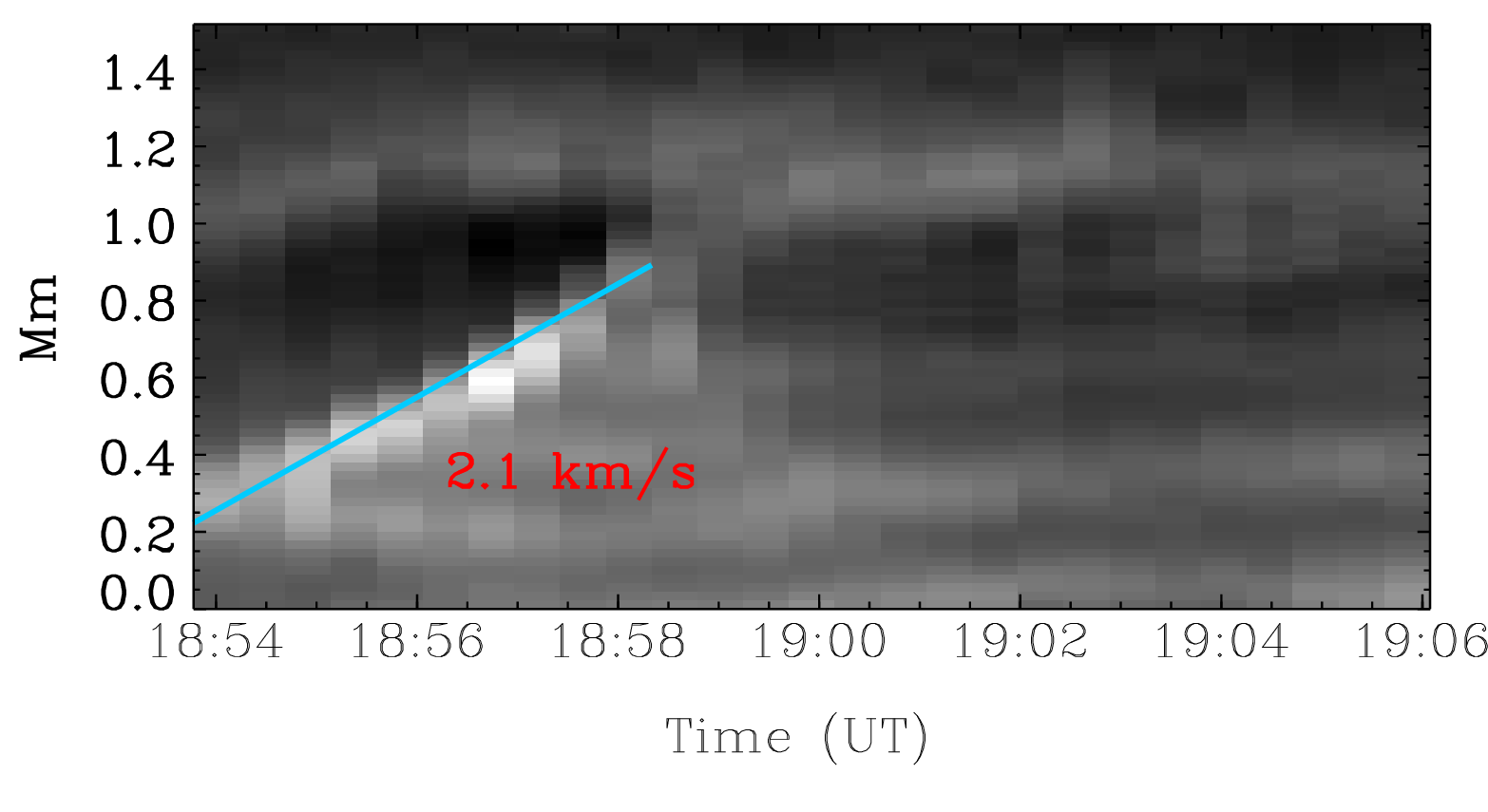}
\end{minipage}

\caption{Time slice diagrams in  two slices  marked in  the $H_\alpha$ blue wing  image at -0.8 \AA{} (left panel). The right panels, from top to bottom,  show  the plasma trajectories moving along the S1 and S2
slices, respectively. S1 indicates the  sweeping motion of the front of the jet and S2 the slipping motion of the bright footpoint of the jet, which  ultimately invades the sunspot umbra.
The footpoint of the jet  moves by  about $2.1~\rm{km\,s}^{-1}$ which is around  $12~\rm{^{\circ} min^{-1}}$.  
 The top of the jet moves with a speed of $6.9~\rm{km\,s}^{-1}$.}
\label{slice_diagram}
\end{figure*}
 \subsection{Dopplergram of the fan-shaped jet}
 
 In order to determine the line-of-sight (LoS) velocity  of the fan-shaped jet material, we created Doppler diagrams of the plasma using the $H_\alpha$ 11 wavelength images  from the GST observations. We calculated the center of weight of the $H_\alpha$ line profile at each pixel to estimate the Doppler shift relative to the reference line center. We averaged the entire observing FOV to obtain the reference line center (except the region where the sunspot is located) and all the line profiles were corrected by comparing them with a standard $H_\alpha$ profile, obtained from the NSO/Kitt Peak FTS data \citep{Su2016}.

\begin{figure*}[!htbp]
\begin{minipage}{\textwidth}
\centering
\includegraphics[width=80mm,angle=0,clip]{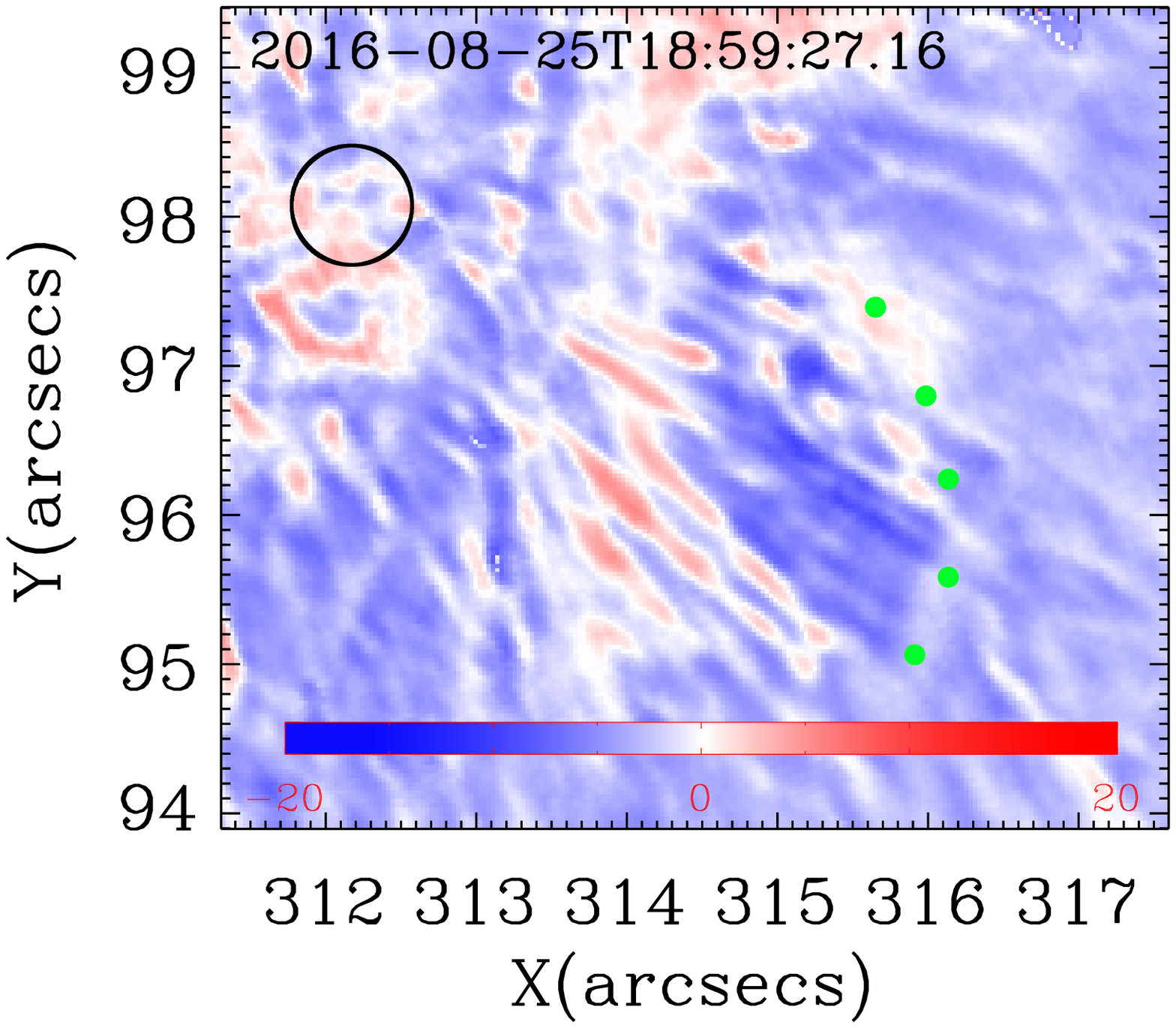}
\includegraphics[width=80mm,angle=0,clip]{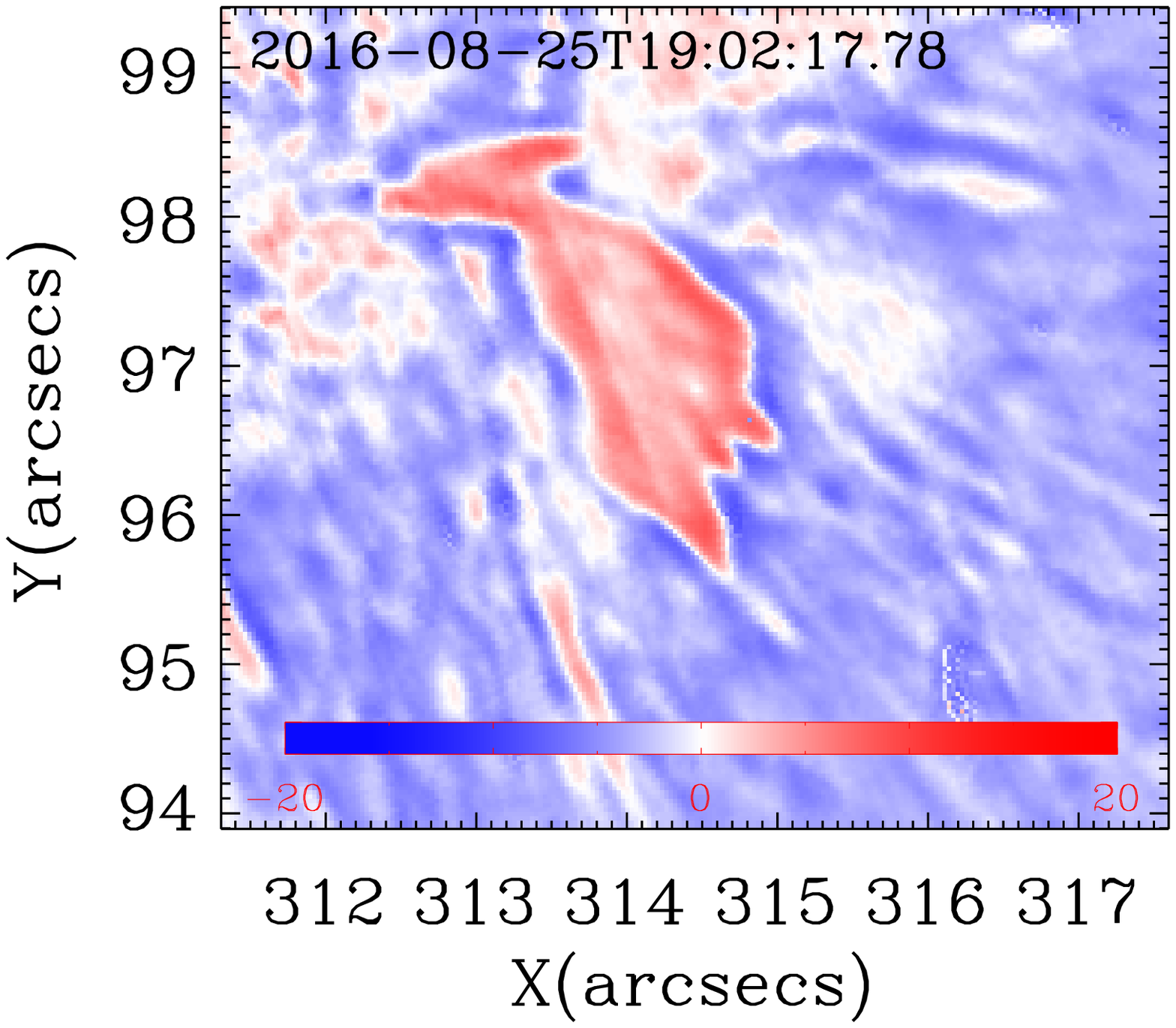}
\end{minipage}
\caption{LOS Doppler velocity maps obtained from GST/VIS of the  zoom-in area of R1 (same field of view as in Figure \ref{Ha_lineprofile} bottom left). The footpoint and front of the jet are marked by a  black circle and green dots, respectively.}
\label{doppler_vel}
\end{figure*}

  Dopplershift maps presenting  the LoS velocities are shown in  Figure~\ref{doppler_vel}  with  blue and red colors  for the blue and redshift motions.
 Before 19:00:21 UT, the Doppler blueshift of the jet is dominant   indicating plasma flowing upward, after that time   the redshift is leading, indicating downward flows. According to the calculation of  LOS velocities, we found that the upflow speed is up to $-14.1~\rm{km\,s}^{-1}$ and  the downflow  speed of the order of $9.5~\rm{km\,s}^{-1}$.  Plasma fell back toward the chromosphere and four minutes later, the jet burst was complete. 

\subsection{Footpoint of the jet}
\label{footpoint of the jet}

 At 18:52:54 UT, a small bright point is identified  in the $H_\alpha$ red wing at + 0.8 \AA{} at the jet footpoint  $[x,y]=[312.9^{\prime\prime},97^{\prime\prime}]$ highlighted by a red circle in Figure 1, panel c1. Between 18:52:54 UT and 18:58:04 UT the intensity of this bright point increases and it moves northward up to $[x,y]=[312.8^{\prime\prime},97.6^{\prime\prime}]$ before totally disappearing  at 18:58:04 UT. With a height-time diagram, we can determine  the bright point velocity  on  the order of $2 ~\rm{km\,s}^{-1}$ (Figure 2, bottom-right panel).

\begin{figure*}[!htbp]
\begin{minipage}{0.5\textwidth}
\centering
\includegraphics[width=90mm,angle=0,clip]{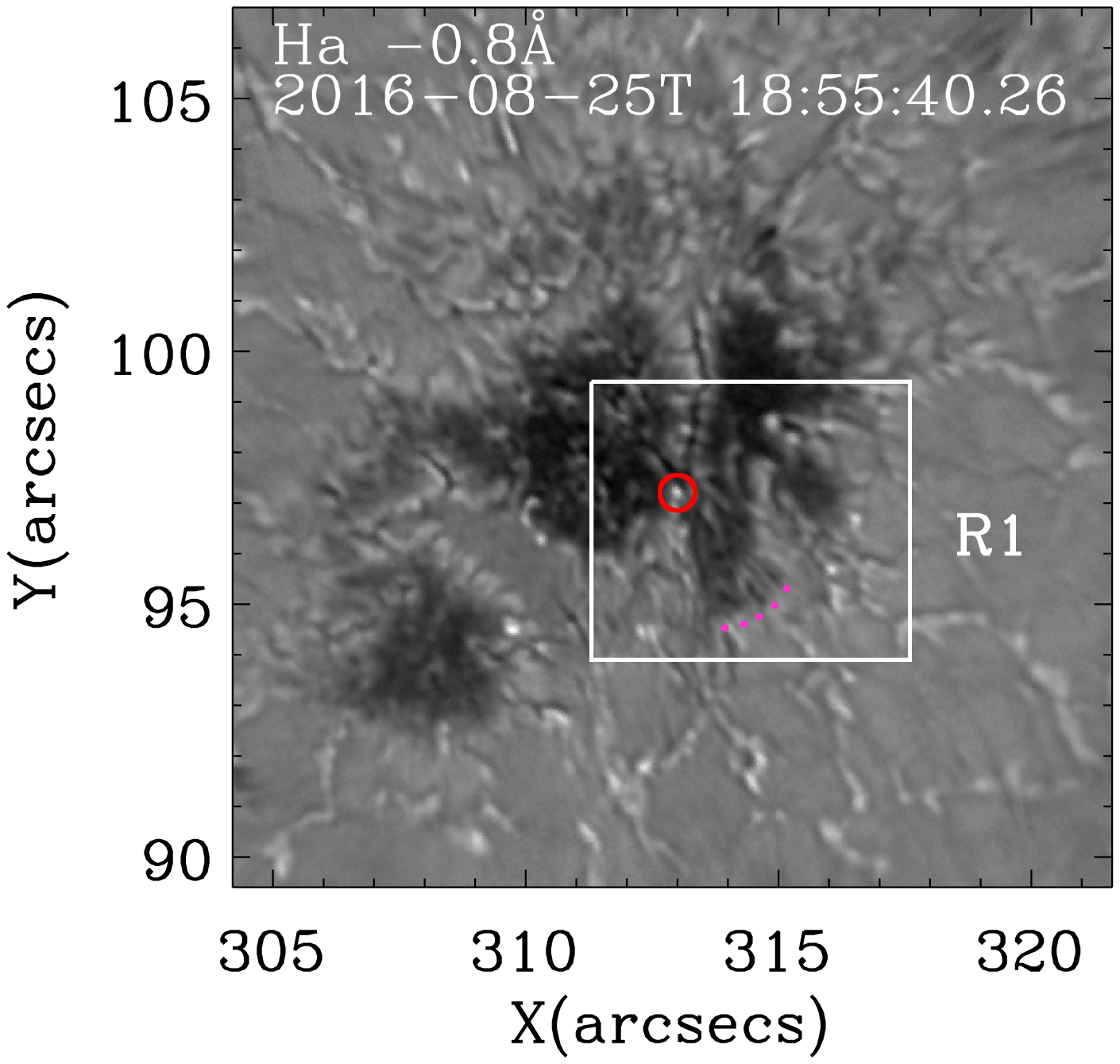}
\includegraphics[width=90mm,angle=0,clip]{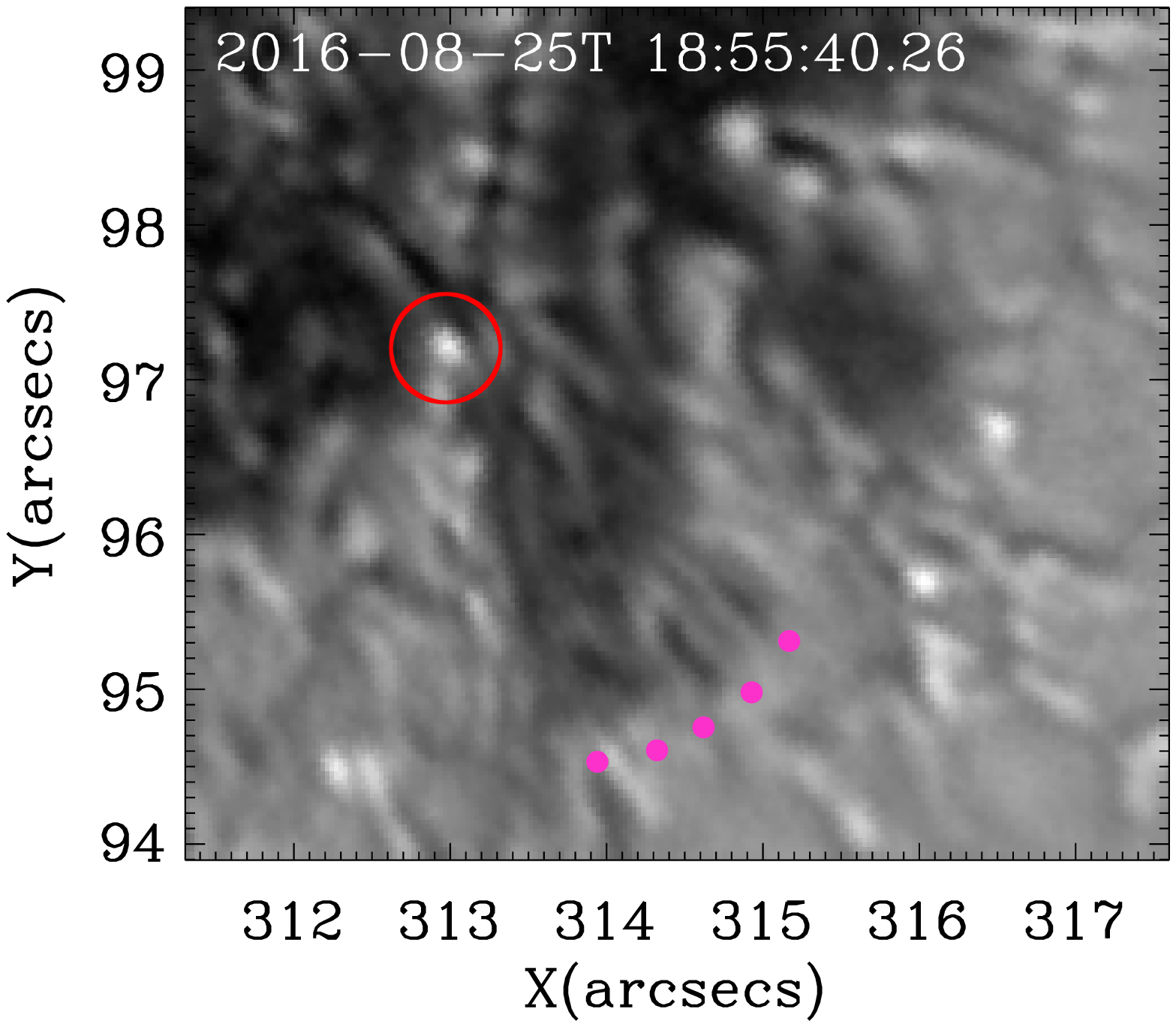}
\end{minipage}
\begin{minipage}{0.5\textwidth}
\centering
\includegraphics[width=90mm,angle=0,clip]{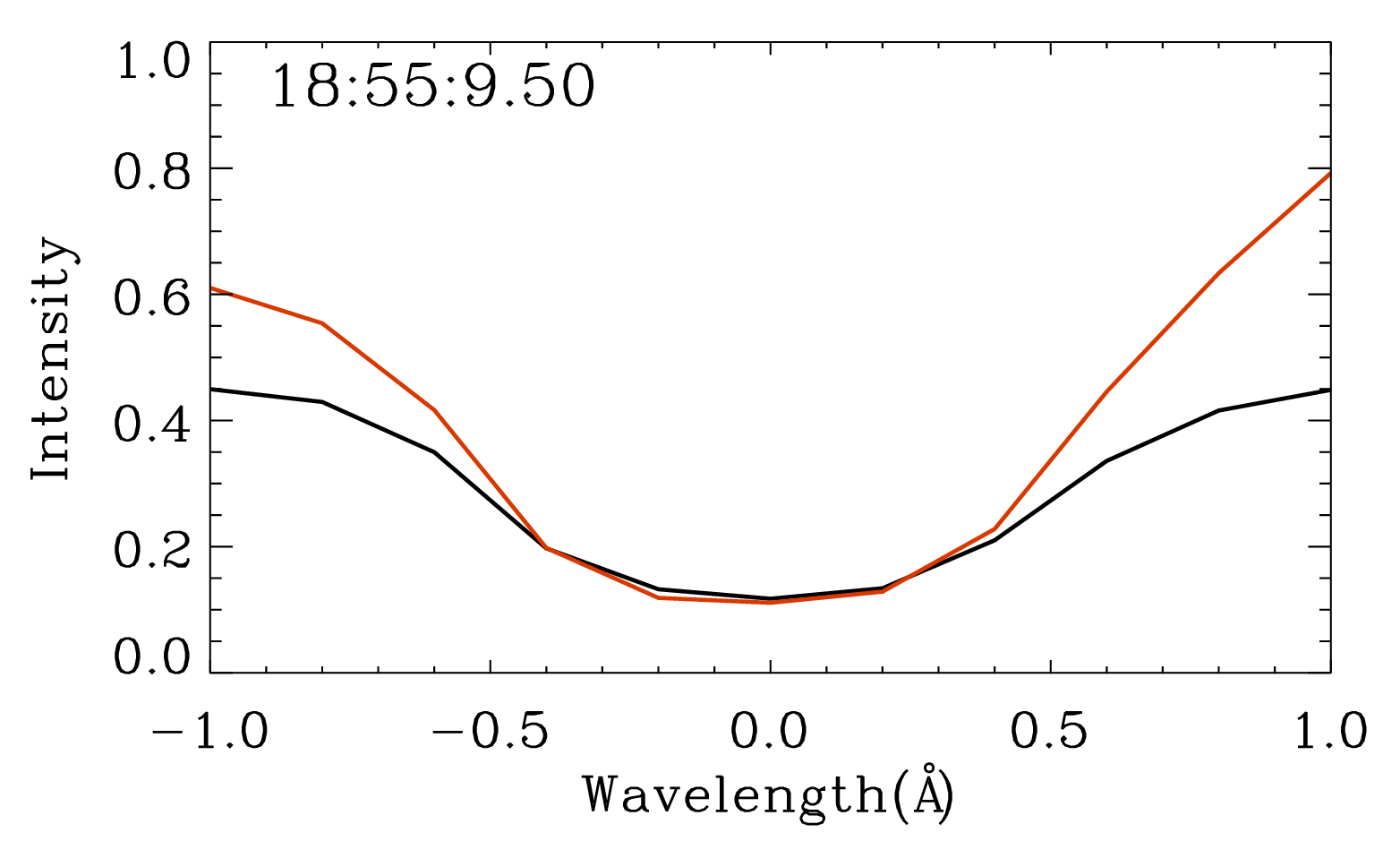}

\includegraphics[width=90mm,angle=0,clip]{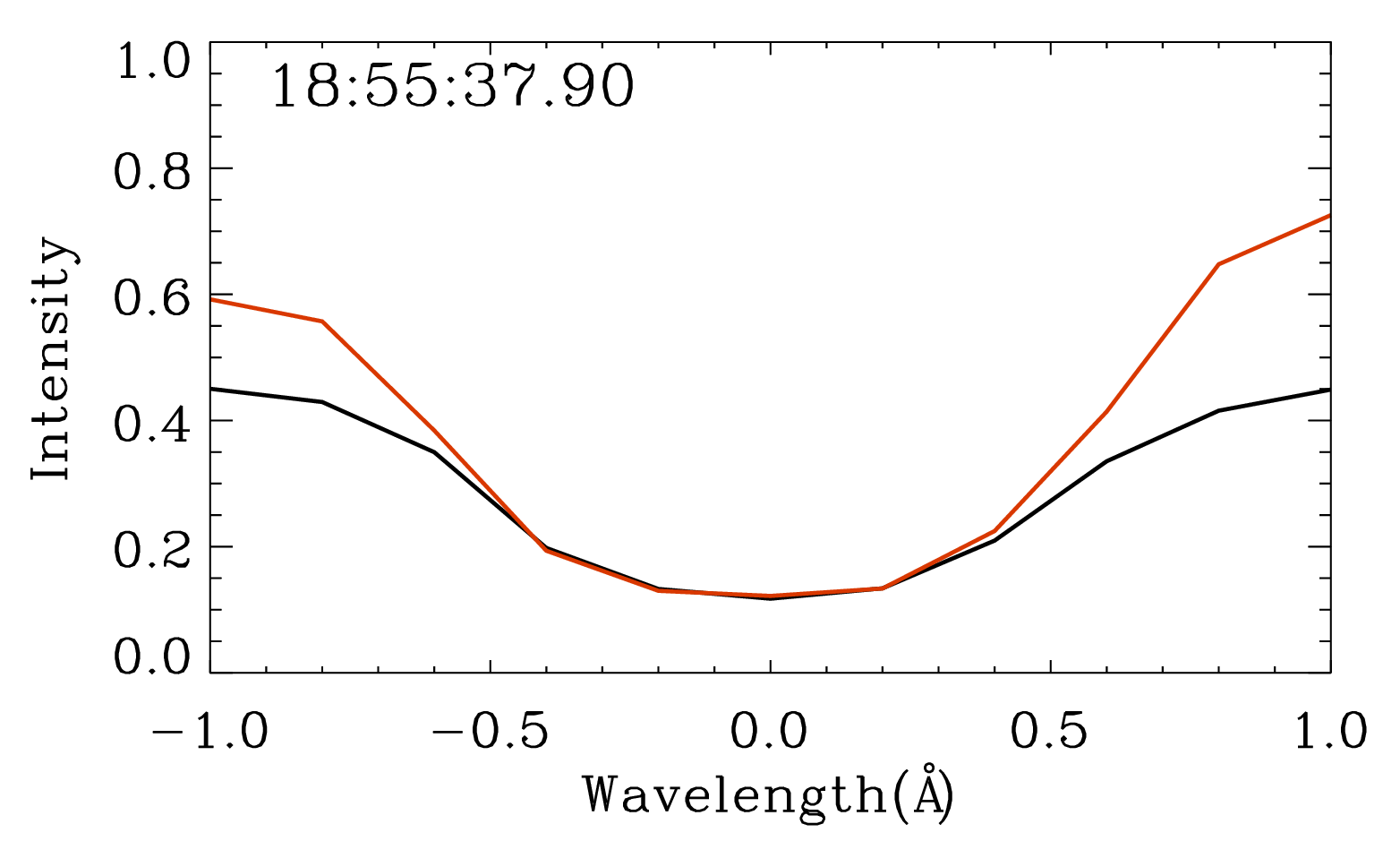}

\includegraphics[width=90mm,angle=0,clip]{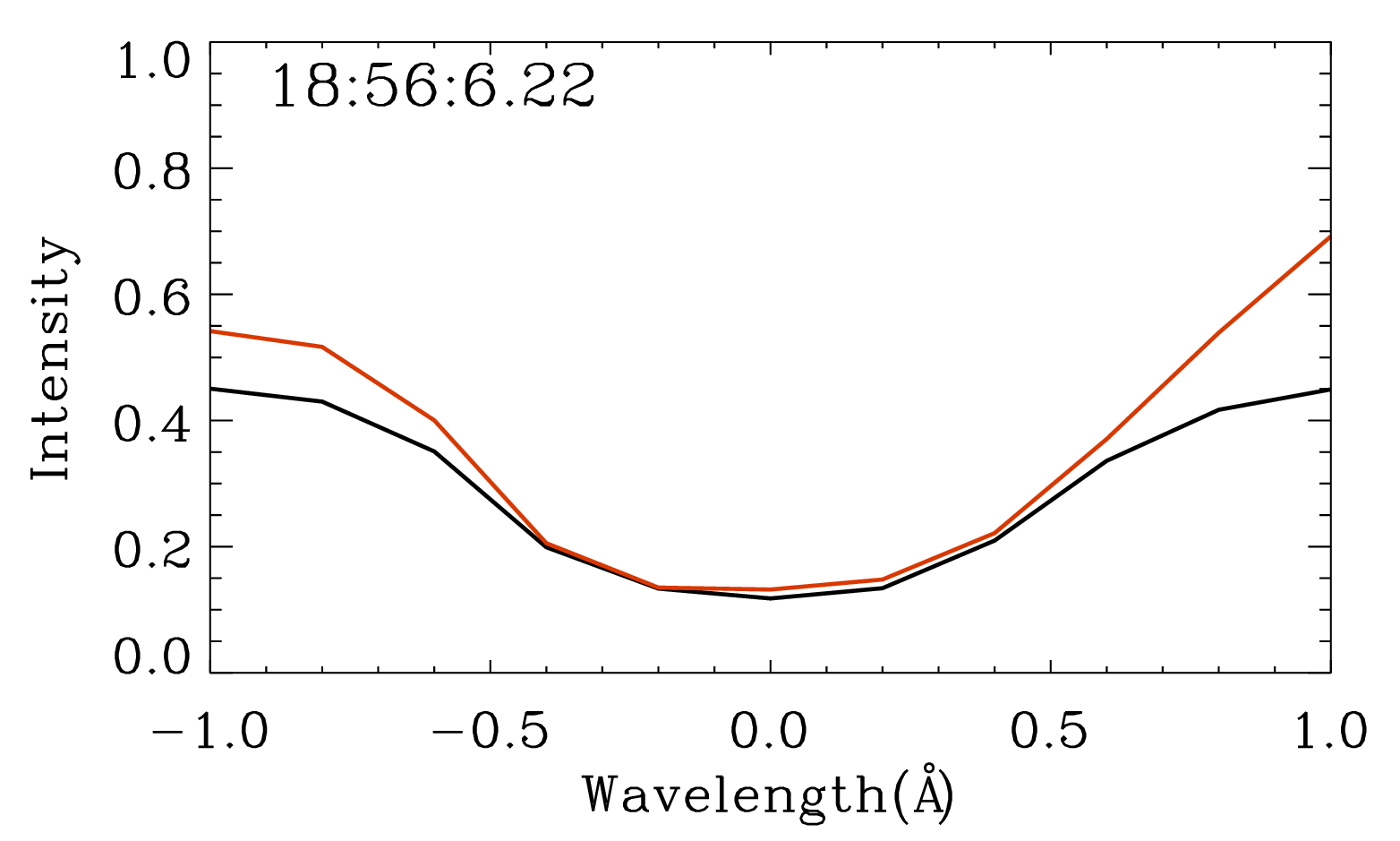}
\end{minipage}

\caption{Footpoint (red circle) and front of the jet  (magenta dots) are marked in R1 region and  in a zoom-in area of R1 (left panels).  Normalized $H_\alpha$ spectral profile (red line) at the footpoint of the jet  and reference profile (black line) for three times during  magnetic reconnection are shown in the right
panels.}
\label{Ha_lineprofile}
\end{figure*}

We analyzed the characteristics of the $H_\alpha$ spectral lines at the  jet footpoint. We normalized the intensity and exposure time of the  $H_\alpha$  11 wavelength images.   The $H_\alpha$ profiles in the bright points  corresponding to the jet footpoints are presented  at three different times at 18:55:09, 18:55:57, 18:56:06 UT in Figure \ref{Ha_lineprofile}. They  show a  clear  enhancement of  intensity in the $H_\alpha$  line   wings  and  no obvious signature in the $H_\alpha$  line core. These profiles are  consistent with Ellerman bomb (EB) $H_\alpha$ line profiles  \citep{Pariat2007,Hashimoto2010,Watanabe2011,Rouppe2016}.  It is worth noting that the reference $H_\alpha$ profile (averaged spectral profile of $H_\alpha$ on the quiet region) and the bright point profiles are symmetric. The intensity in the blue wing is very close to the one in the red wing which is typical of  EB$'$s profiles.  Ellerman Bombs can be explained as the radiative signatures of magnetic reconnection episodes occurring in the photosphere \citep{Chen2001,Fang2006,Grubecka2016,Joshi_J2020}. Therefore, observing EB$'$s typical signature at the footpoints of the fan-shaped jet during the first phase, may indicate that the observed fan-shaped jet could have been driven by magnetic reconnection.

The active region is in a decaying phase with a  very fragmented leading sunspot due to  magneto-convection as it is shown in  the TiO and $H_\alpha$ wing images (see Figure 1 and the corresponding movie). In  the TiO and H$\alpha$ GST images (Figure 1), the LB is clearly identified extending from  $[x,y] = [311.3^{\prime\prime},95.8^{\prime\prime}]$ to  $[x,y] = [313.8^{\prime\prime},101.2^{\prime\prime}]$.  The bright point is moving along the LB axis. According to \citet{Tian2018,Lim2020}, LBs can be the signature of magnetic flux emergence inside a pre-existing sunspot. Therefore, magnetic reconnection may occur between the emerging flux and the sunspot magnetic field. The presence of a LB is an additional information supporting the fact that the bright point results from magnetic reconnection that could trigger the ejection of chromospheric material observed as the fan-shaped jet.

 \subsection{Photospheric magnetic field}
 \label{the photospheric magnetic field}
 
\begin{figure*}[!htbp]
\begin{minipage}{\textwidth}
\centering
\includegraphics[width=80mm,angle=0,clip]{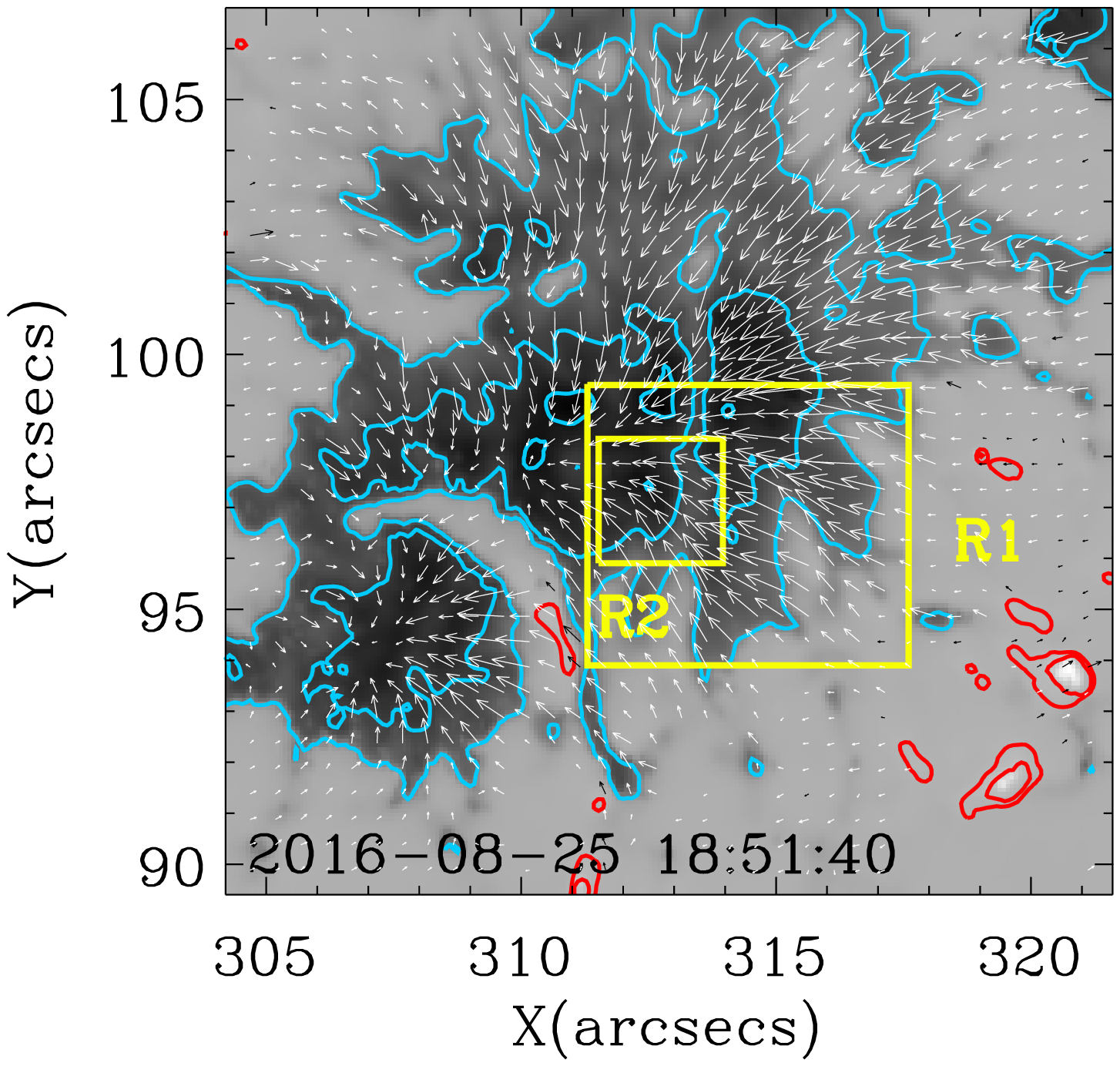}
\includegraphics[width=80mm,angle=0,clip]{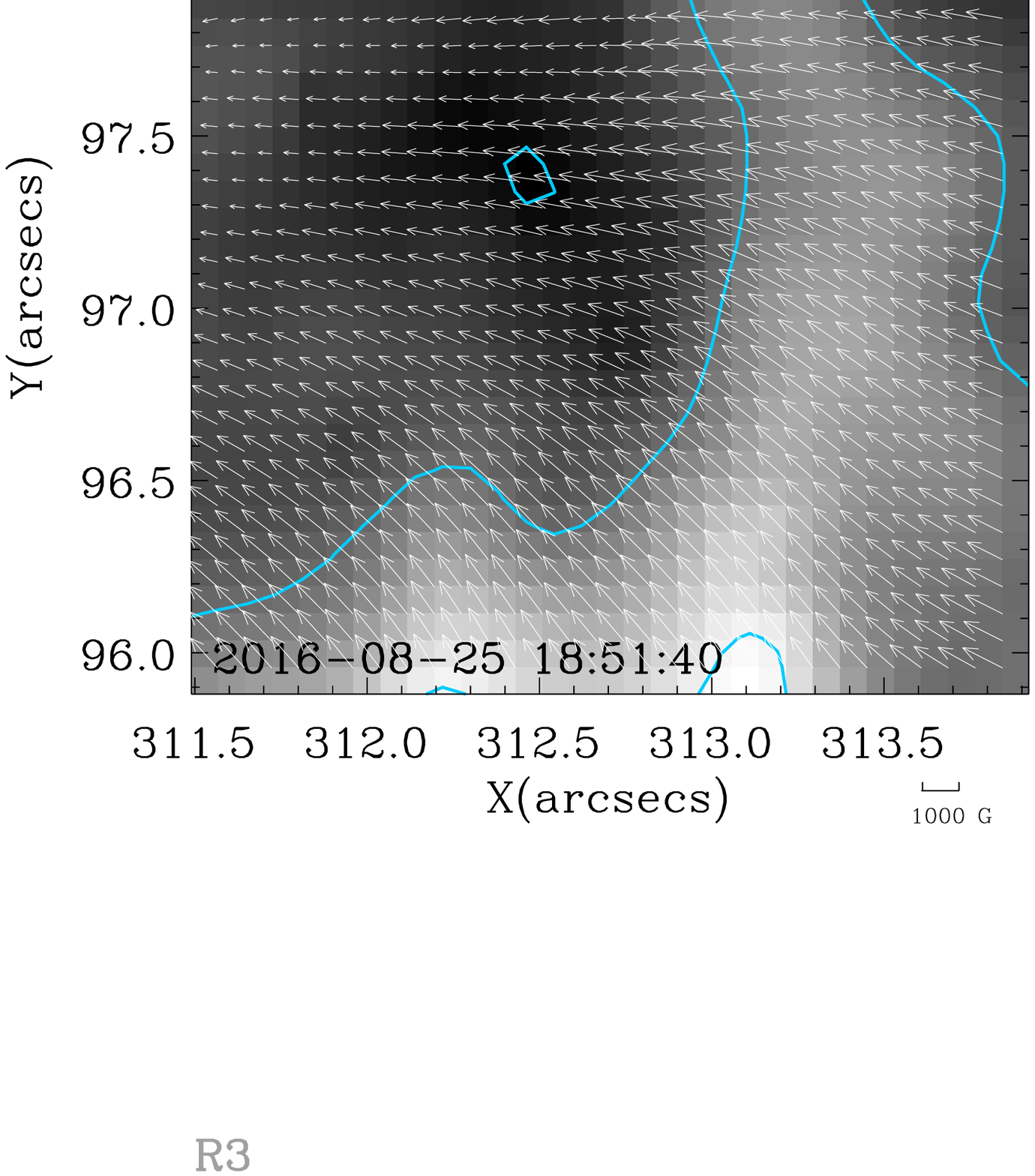}
\end{minipage}
\caption{Vector magnetogram of the  FOV and a zoom-in area of R1 obtained from GST/NIRIS after removing the $180~\rm{^{\circ}}$  ambiguity in the transverse field. The blue and red contours indicate the magnetic field Bz  at -2300G, -1700G, -900G, 50G, and 150G respectively. The vertical component of vector magnetic field Bz in grey scale is overlaid with arrows.
The arrows indicate the strength and direction of the transverse magnetic field. } 
\label{vector_magnetogram}
\end{figure*}

Figure~\ref{vector_magnetogram} left panel shows the NIRIS vector magnetogram  of the sunspot  at the time of 18:51:40 UT obtained by applying the  Milne Eddington (ME) inversion to the  
Stokes profiles of FeI 1565~nm doublet using the inversion code of J. Chae \citep{Landi1992}. The azimuth component of the inverted vector magnetic field was processed to remove the $180 ^{\circ}$ ambiguity \citep{Leka2009}.  

We find that the  identified LB has a weaker vertical magnetic field  (around 1000 Gauss) compared to the surrounding umbra  (1700 Gauss)  and the  magnetic field in its upper part (from  $[x,y]=[311.3^{\prime\prime},94^{\prime\prime}]$ to $[x,y]=[317.6^{\prime\prime},99.3^{\prime\prime}]$.
 The horizontal component of the magnetic field reaches also 1000 Gauss in the LB. The arrows  
in Figure \ref{vector_magnetogram} (R1 box)  indicate the direction of this component. They are  
highly inclined  versus the axis of the LB close to the y  axis with a tendency of being aligned with the main axis of the LB.  This is in agreement with the description of the magnetic field in LBs \citep{Leka1997}. However, we did not find a pure horizontal field (as would be expected for a LB) and we did not detect any opposite polarity  in the area of the bright point which could  lead to a jet-type reconnection as proposed in different models of fan-shaped jet in LBs  \citep{Bharti2015,cheung2015,Robustini2018}. 

\begin{figure*}[!htbp]
\begin{minipage}{\textwidth}
\centering
\includegraphics[width=150mm,angle=0,clip]{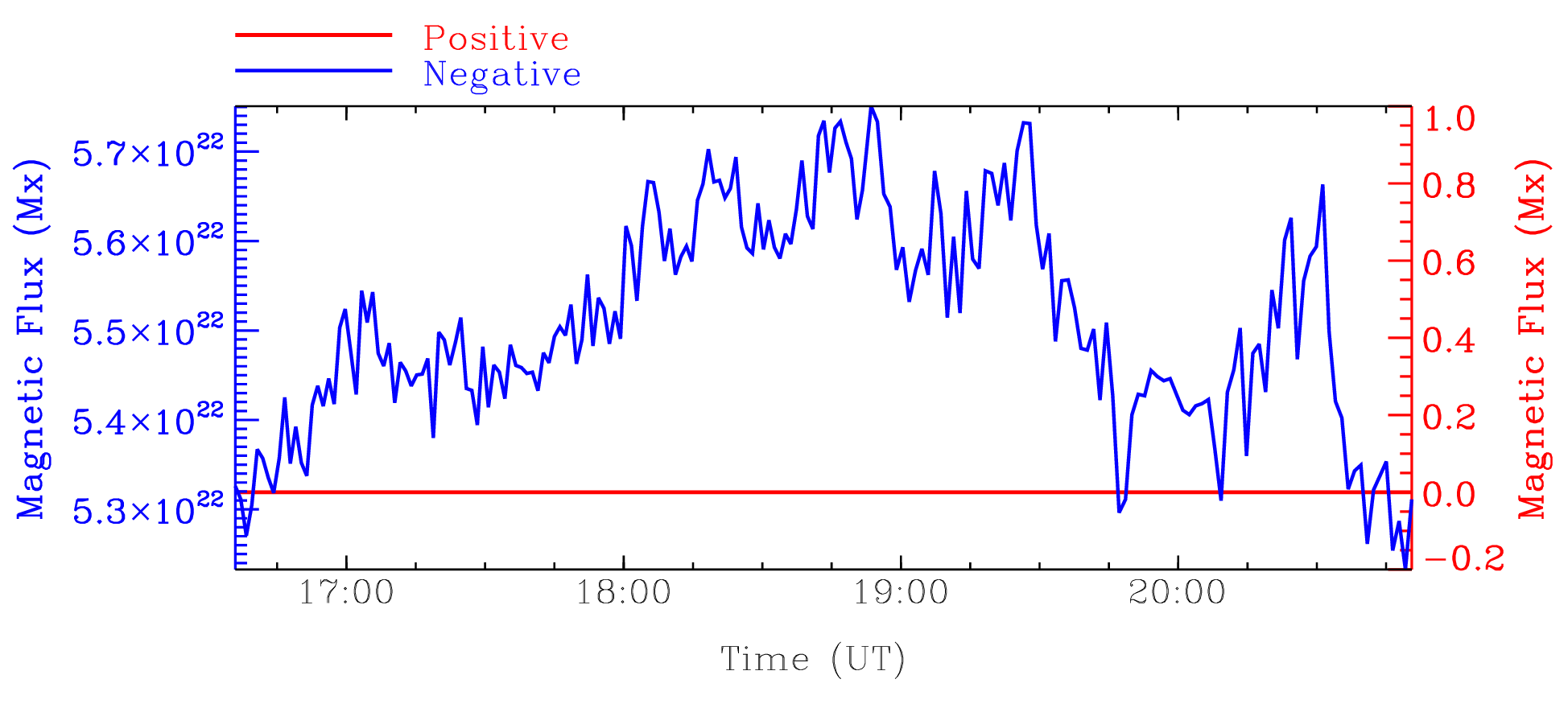}
\end{minipage}
\caption{Variation  of  both positive and negative magnetic flux in absolute values of the  longitudinal magnetic field of R2 region marked in Figure \ref{vector_magnetogram}. During the jet eruption, some new flux emerges and magnetic reconnection may occur. The evolution of negative and positive flux are represented by blue and red color, respectively.}
\label{magnetic_flux}
\end{figure*}

By computing the magnetic flux in the region R2, we found no emergence of new positive magnetic flux  (see Fig. \ref{magnetic_flux} on the flux in R2).   Pixel by pixel, we analyzed   the Stokes profiles to detect a third lobe but we could not identify such a pattern in the profiles, as done in the work by \citet{Bai2019}.  The difference of resolution of NIRIS compared to VIS is only a factor 2.8 and this cannot explain the non-visibility of an opposite polarity either.

The right panel in Figure \ref{vector_magnetogram}  displays a zoom on the R2 box of the left panel. It shows the south part of the LB where the bright point associated with the fan-shaped jet footpoint moves during the event. We observe a change in direction versus the x axis of the horizontal component evolving from a south-west to north-eastward orientation to a west-eastward  orientation in the northern part. This change in the 
horizontal field direction suggests a high gradient of connectivity at the footpoint of the fan-shaped jet.

This  change in the 
horizontal magnetic field  direction is kept as we go further away from the fan-shaped jet footpoint toward the west (see the right side of the R1 box). This change in orientation in  the field is consistent with the diverging shape of the dark structures forming the fan-shaped jet observed in the $H_\alpha$ -0.8~\AA{}.


\subsection{Brightenings at the jet front}

  In the course of the fan-shaped jet  event (Section~\ref{fan-shaped jet}), we also identified EUV brigthenings observed in the AIA channels.  The left column of Figure 7 shows  the AIA images at 171~\AA{} for three instances. The red circle and the pink dots show respectively the location of the fan-shaped jet bright footpoint (section~\ref{footpoint of the jet}) and the front of the fan shaped-jet in the $H_\alpha$ blue wing (section~\ref{fan-shaped jet}). While no 171~\AA{}  brightening is associated with the fan-shaped jet footpoint, there are 171~\AA{} brightenings located at the front of the fan-shaped jet. Over the time interval of the fan-shaped jet, these EUV brightenings have a motion that is similar to the one of the fan-shaped jet along the south-east and north-west axes (Section~\ref{fan-shaped jet}).
 
  These brightenings at the front of the fan-shaped jet are observed in all AIA channels except at 304~\AA{}.
  This can be explained  by the fact that the 304~\AA{}  filter is dominated by the emission of the He II  line at 303.78~\AA,{} which is an optically thick line formed at chromospheric and transition region temperatures. Therefore, the filament and surge commonly appear dark in this filter. In the other channels, we observed a  few bright pixels  in the front of the jet. Due to the low resolution of AIA compared with the high resolution of the GST, these brightenings do not show any fine structure.
   They appear  blurry in the images  because the pixel size of AIA is large compared to the GST pixel size and, certainly, the intensity  peaks only in a small area, as compared to the AIA pixel area.   However, in the movie, we can see that  they  are moving versus time around the front  as the fan develops. Behind the front, dark absorption features correspond to the cool jet. The brightening is relatively obvious in 131~\AA{}, 171~\AA{}, 193~\AA{}, indicating that the jet  front  contains hot material, up to $10^{5}$ or even $10^{6}$ K.   The brightenings in the front of the jets  may also correspond to enhancement of  density due to compression in the corona.

 We estimate the speed of the EUV brightening at 171~\AA{}. In Figure 6 (left column), the white and light blue dashed lines shows respectivelely the initial and final location of the EUV brightening. Assuming a uniform speed between those two times, we obtain a brightening velocity of approximately $6.61~\rm{km\,s}^{-1}$. This velocity is of the same order of magnitude as the  speed of the fan-shaped jet (section~\ref{fan-shaped jet}).

\begin{figure*}[!htbp]
\begin{minipage}{\textwidth}
\centering
\includegraphics[width=70mm,angle=0,clip]{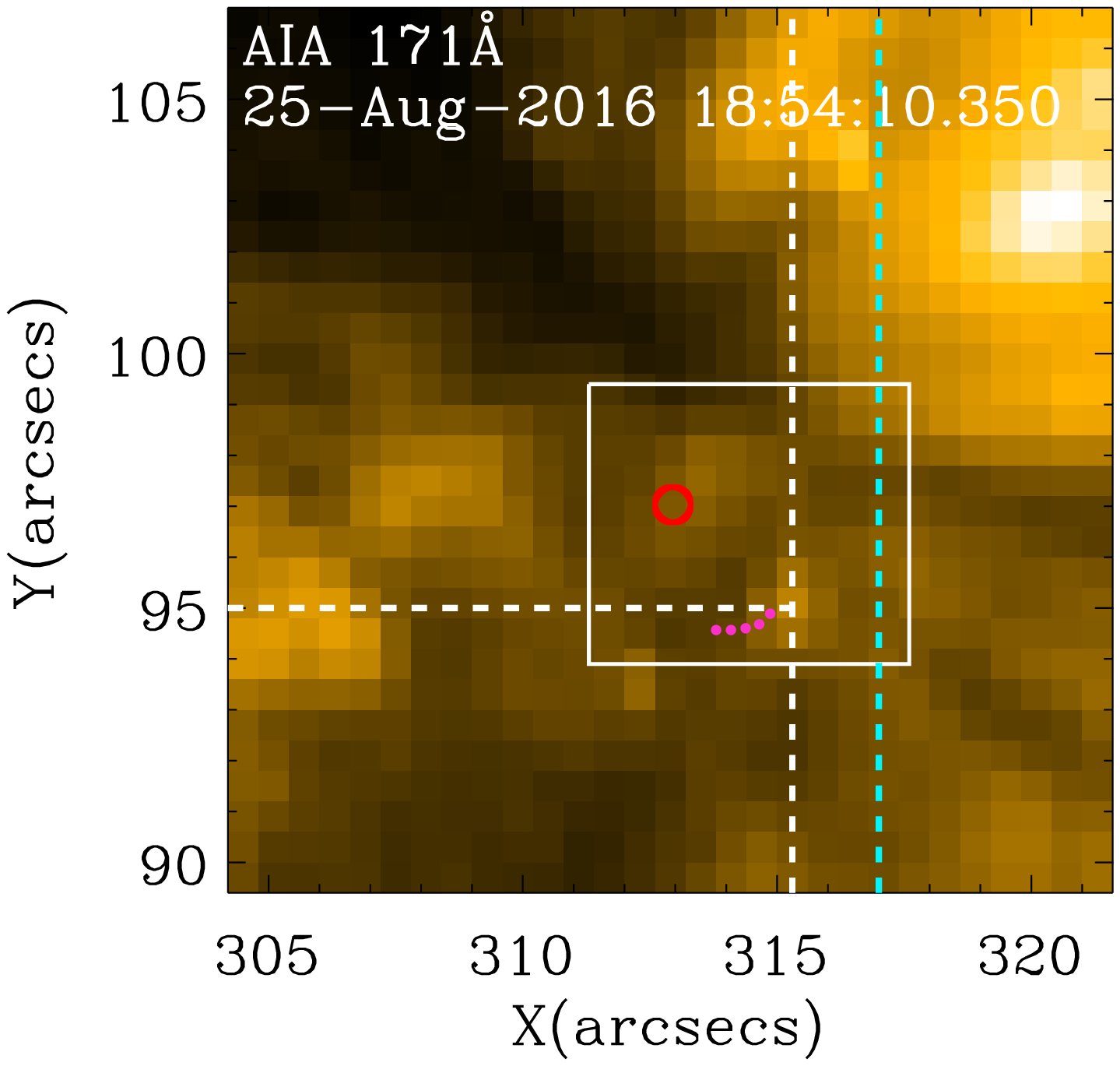}
\includegraphics[width=70mm,angle=0,clip]{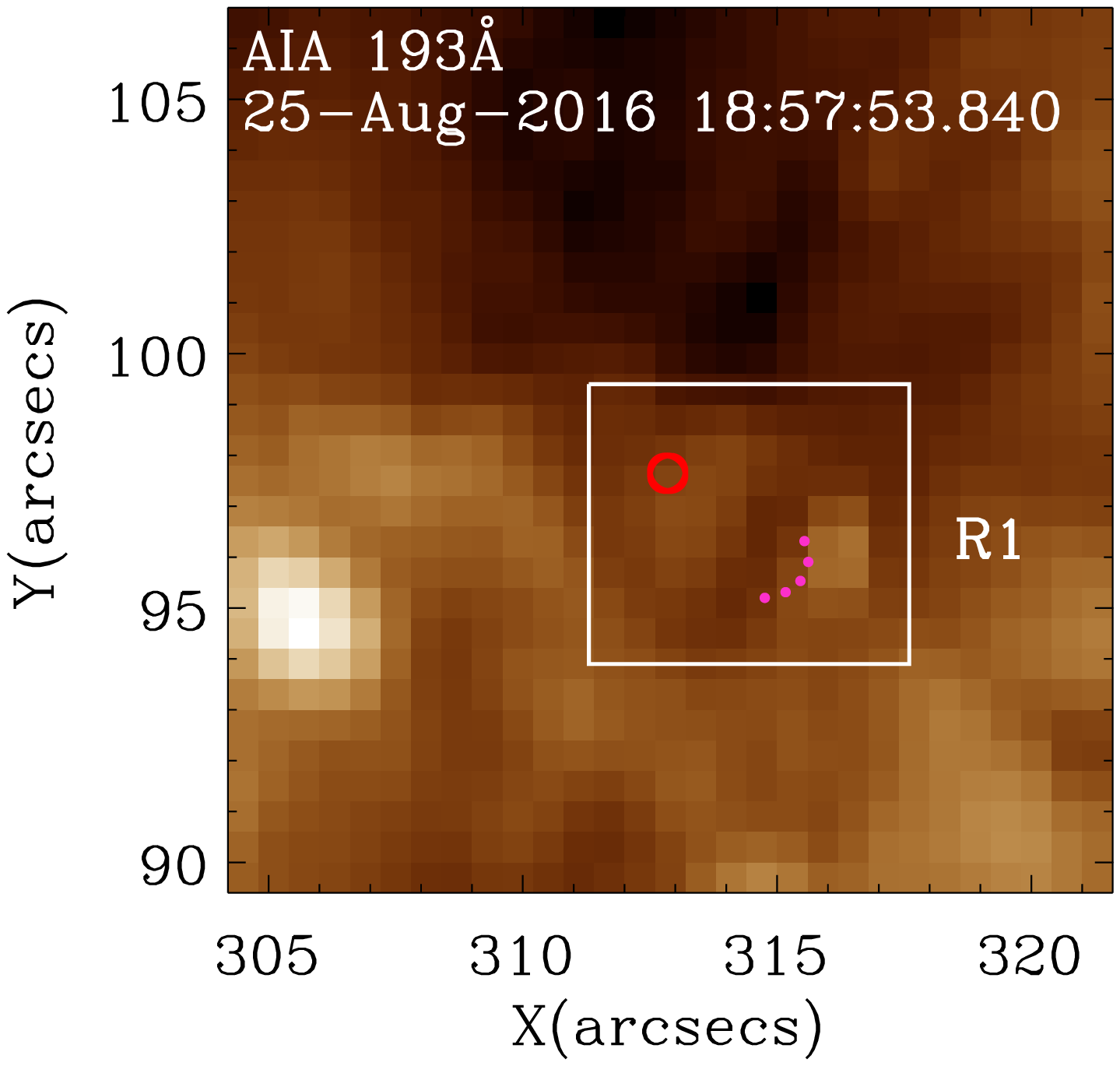}
\includegraphics[width=70mm,angle=0,clip]{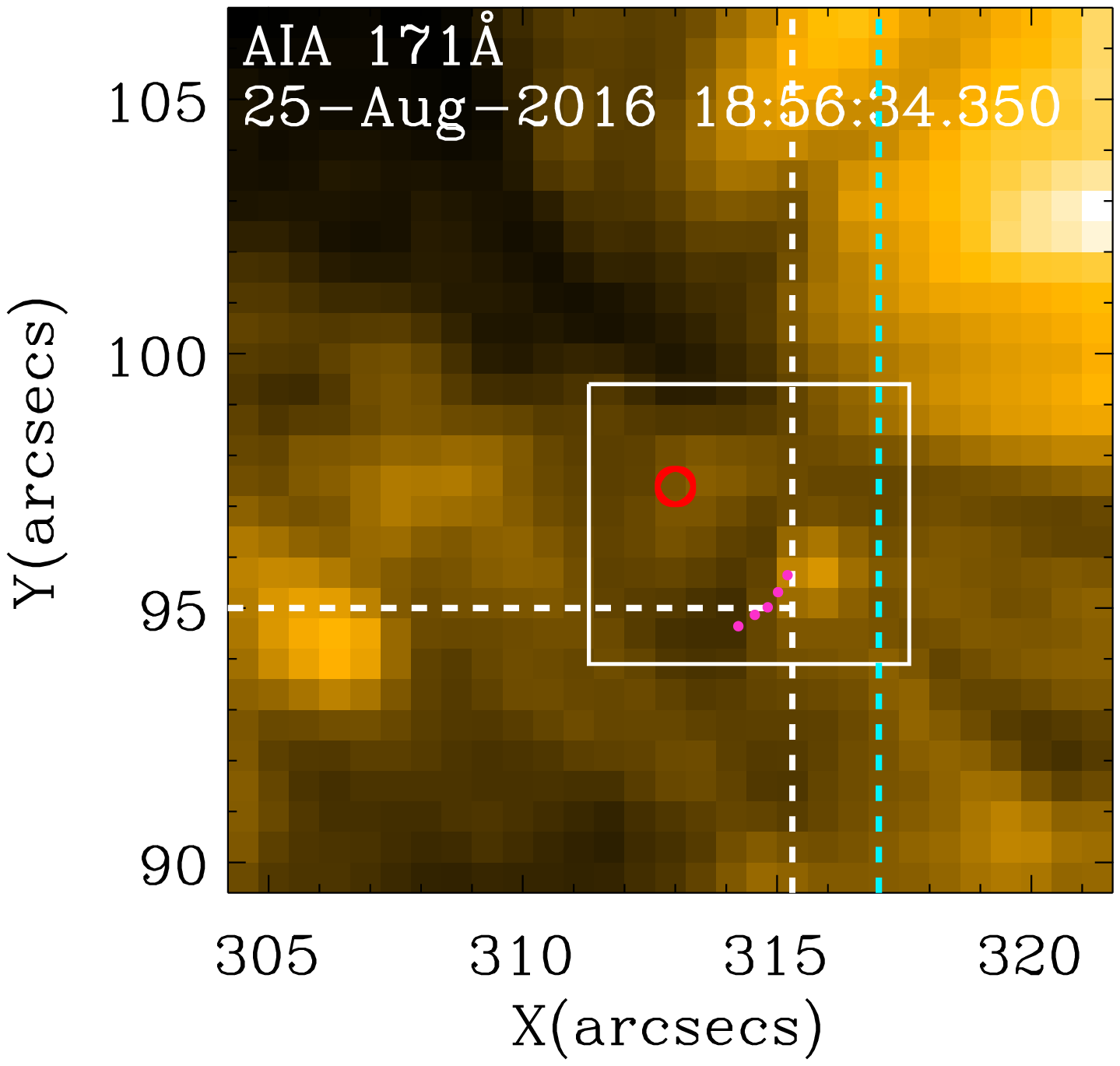}
\includegraphics[width=70mm,angle=0,clip]{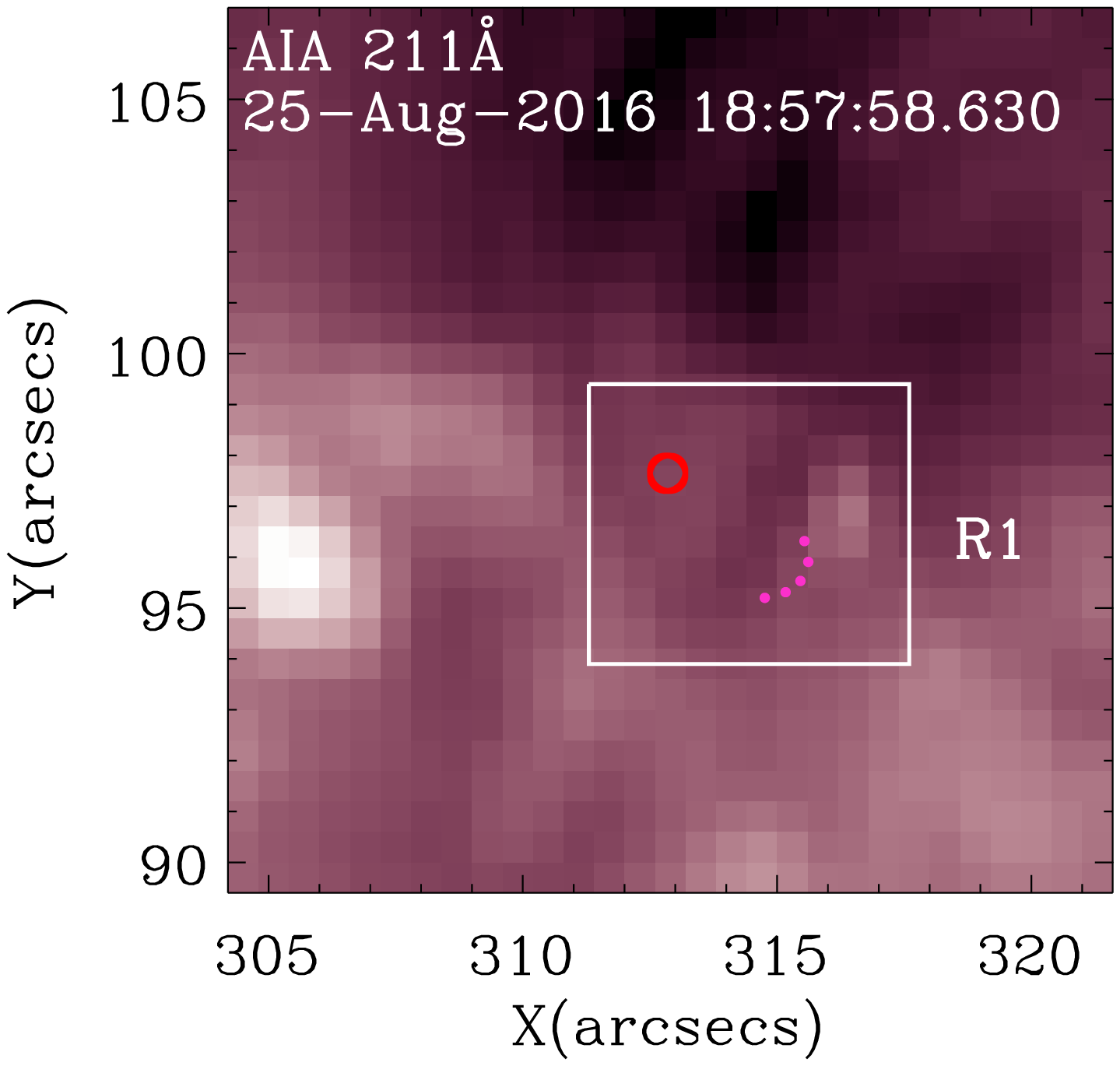}
\includegraphics[width=70mm,angle=0,clip]{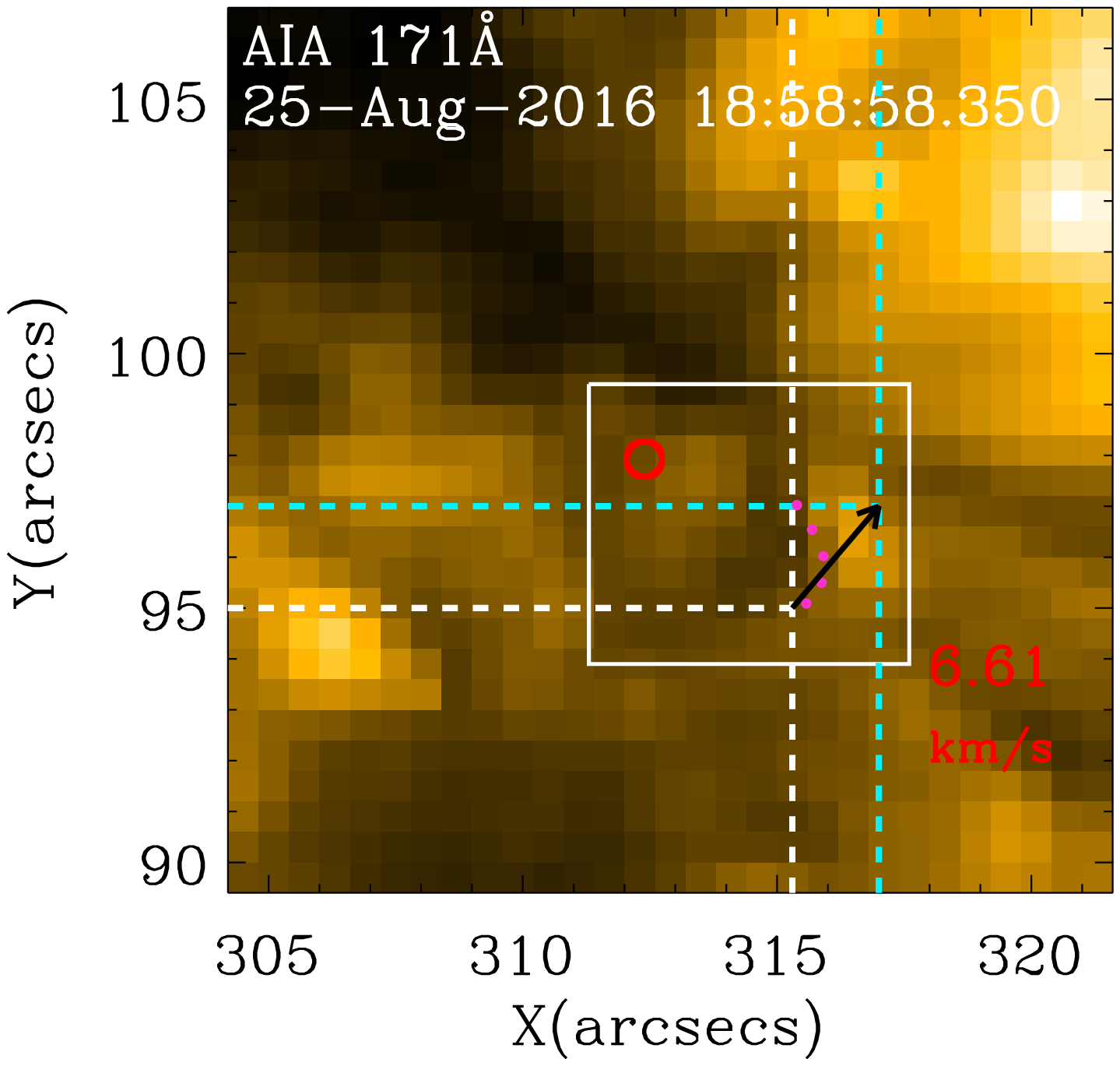}
\includegraphics[width=70mm,angle=0,clip]{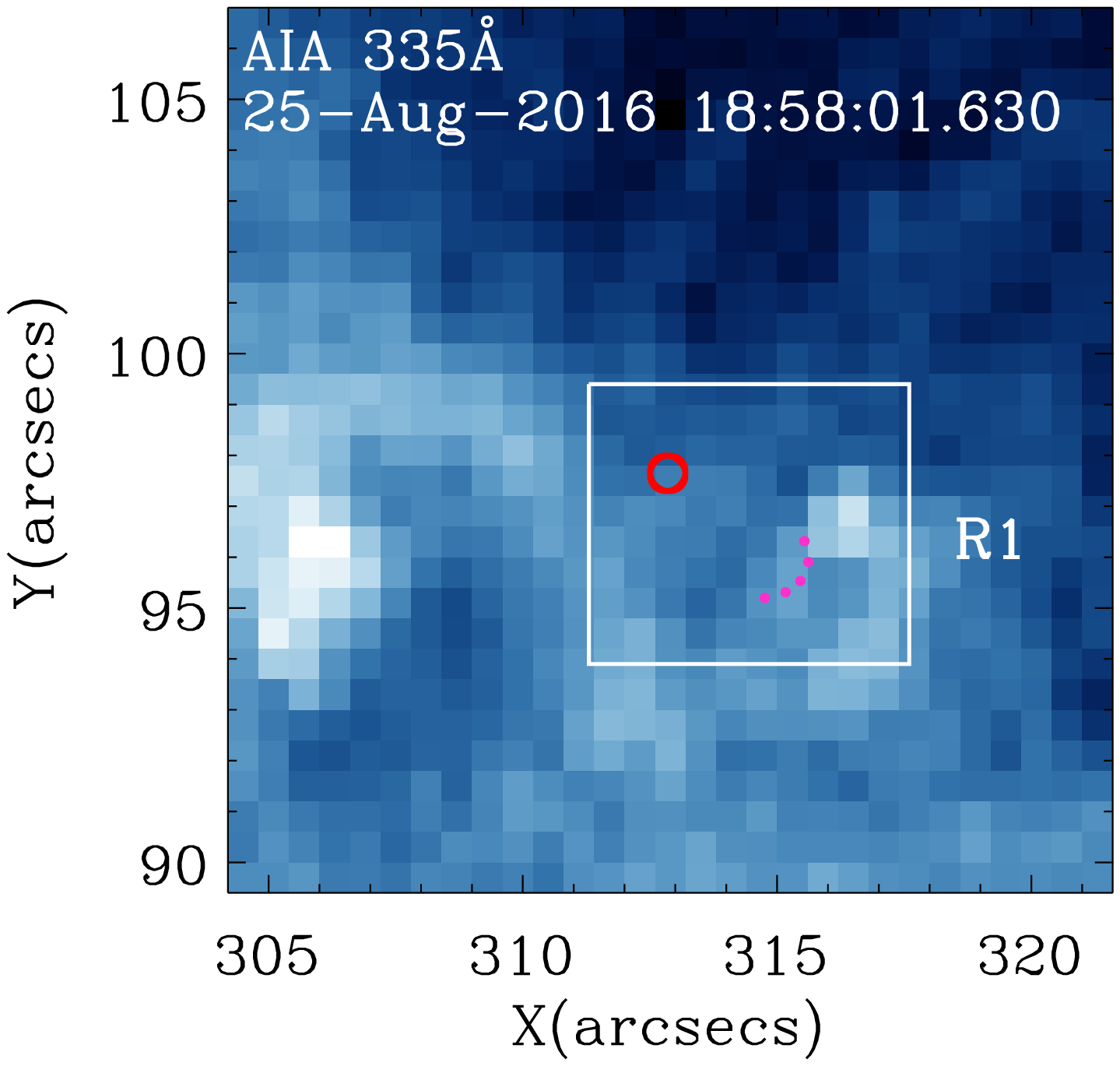}
\end{minipage}
\caption{Corona response observed with the different filters of SDO/AIA. R1 represents the area where the jet occurred. The footpoint and front of the jet are marked by red circle and magenta dots, respectively, as in Figure \ref{images_BBSO}. The brightening  displacement and speed are marked on the bottom left corner panel with an arrow and a value of 6.61 km/s. }
\label{image_AIA}
\end{figure*}


\section{Discussion and conclusions}
In this paper, we report on coordinated observations with the GST at BBSO and SDO/AIA of the active region NOAA 12579 on August 25, 2016. The observed event shows a fan-shaped jet occurring on the  side of a light bridge (LB).

 We obtained the following results:
\begin{enumerate}
 \item The fan-shaped jet observed in absorption in the $H_\alpha$ +/- 0.8~\AA{} shows a diverging shape. In the first phase, it exhibits a sweeping motion from the south to the north of the sunspot with an approximate velocity of $3.15-6.91~\rm{km\,s}^{-1}$ and in the the second phase, the jet material is flowing back toward the photosphere. The computed Doppler diagram of the jet confirms these upflows and downflows (Section~\ref{fan-shaped jet}). 
 
 \item In the GST  observations in TiO, we identified  the  LB  close to the jet that divides the sunspot into two parts. However, the magnetic field analysis does not show any emergence of opposite polarity, as has been theoretically  predicted for LBs (Sections \ref{footpoint of the jet} and \ref{the photospheric magnetic field}). 
 
 \item We also identified a   bright point in the $H_\alpha$ blue wing located in the LB  that is spatially consistent with  the footpoint of the fan-shaped jet. This bright point is moving along the  LB  axis with an estimated speed of $2.1~\rm{km\,s}^{-1}$. By analyzing the $H_\alpha$ lines, we found that this bright point has the same  spectral characteristic as Ellerman Bombs (section~\ref{footpoint of the jet}). This suggests that this bright point may result from magnetic reconnection.

 \item  In analyzing the NIRIS  vector magnetogram, we found that the magnetic field associated with the fan-shaped jet strongly diverges starting at the jet footpoint up to its front.  

\item  From the AIA observations, during the first phase of the jet, we observe a multi-thermal structure located at the front of the $H_\alpha$ jet and moving in time with it from south to north with an estimated speed of $6.6~\rm{km\,s}^{-1} $. This indicates that hot material up to  $10^6~\rm{K}$ is present at the  fan-shaped jet front.

\end{enumerate}
  
     According to our analysis, we propose that the fan-shaped jet corresponds to the ejection of material along reconnecting magnetic field lines. The magnetic reconnection occurring in the photosphere-chromosphere is  supported
     by the observation of EB$'$s typical radiative signature at the bright jet$'$s footpoint.

 The diverging magnetic field supports the idea that the fan-shaped jet occurs in Quasi-separatrix layers (QSLs), which are topological elements connecting a bipolar field with a strong magnetic connectivity gradient, where thin and intense current sheet can be created \citep{Demoulin1996}. When 3D magnetic reconnection occurs in QSLs it creates an apparent slipping motion of the reconnecting field lines \citep{Aulanier2006}. Such  a slipping motion of the field lines can be indirectly observed with the plasma filling the reconnecting field lines. In our event, the slipping reconnection may explain the shape and the dynamics of the fan-shaped jet observed in the  $H_\alpha$ line.

Since the fan-shaped jet is anchored in the LB, an alternative scenario could be that the fan-shaped jet anchored in LB is a jet created by magnetic reconnection between the LB$'$s emerging flux and the surrounding magnetic field of the sunspot. However, in our study, we did not identify in the NIRIS  magnetogramm  opposite magnetic field emerging in the LB area (as we would have expected).  We analyzed the HMI vector   magnetograms  during 24 hours before the event and no opposite polarity was detected -- not even  the LB  was resolved. Nonetheless, there is a factor  of 6 (0.5 divided by 0.081 is equal to 6.17. The pixel size of HMI data before pre-processed is $0.^{\prime\prime}5$ and the pixel size of magnetic field data is $0.^{\prime\prime}081$) in the spatial resolution between GST/BBSO and HMI/SDO. It is likely that HMI/SDO  is not equipped with the required resolution to detect emerging bipole  and fast evolving LB. 
  
Another  interesting idea comes from  theoretical models of sunspot formation in a magneto-convection environment  (see Fig. 22 in \citet{Rempel2012,Bharti2020}).  The sunspot, with its  ephemeral LBs, is in a first phase of decay with intense fragmentation. The LBs are observed appearing and disappearing during the day. At the time of the jet, the LB is in  a formation state with mainly bright points which are the footpoints of the jets.  These bright points are brighter than the surrounding environment and, thus, hotter. They may rise by buoyancy from the subsurface. In theoretical models  it is shown that   magnetic field lines can bent due to down-flows  around  convective cells \citep{Bharti2020}. Consequently it creates  hair-pin shape magnetic field lines and opposite polarities may appear. 
The high spatial resolution  of the GST could not  be  enough to resolve such bent magnetic field lines or the hair-pin turns in the subsurface of the Sun.  We may wait for the new generation of ground-based telescopes such as DKIST and GIANT to resolve these fine structures.
  
It is worth mentioning that if magnetic reconnection occurs between some emerging bipoles and the strongly diverging field of the sunspot, magnetic reconnection across the QSLs can still take place and thus explain the shape and sweeping dynamics of the fan-shaped jet.
 
During the first phase, in addition to the $H_\alpha$ fan-shaped jet, a bright front is observed in most of the AIA wavelengths. Such a brightening indicates that the plasma in this region has been heated. Since it is located right at the front of the fan-shaped jet and it moves along with it at a speed that is on the same order of magnitude as the sweeping motion of the fan-shaped jet, we propose that the plasma may be heated by compression as the cold material propagates along the reconnecting field lines. The brightening disappears at the beginning of the second phase when the plasma is flowing back down toward the photosphere. We can speculate here that the jet material encounters a denser region that halt the propagation of the cold material that is then forced to flow back down.
  
  Finally, we may also consider  that the material inside  the jets (considered as flux tubes) could be pushed up by increasing pressure  due to MHD waves and convection. The  absence of opposite polarities is in favor of this mechanism \citep{Hollweg1982,Dey2022}. 
  A shock would be created in the corona, effectively stopping the ejected material, as proposed by \citet{Iijima2015}. This mechanism ought to be further investigated in the future with high-resolution observations.

\begin{acknowledgements}
      Y. Liu and G.P. Ruan acknowledge the support by the NNSFC grant 12173022, U1831107, 11790303 and 11973031. We are grateful to V\'eronique Bommier for the discussion on HMI vector magnetograms, to Jin Chunlan, Chen Yajie, Hou Zhenyong for the data discussion and  to Vasilis Archontis and Robert Cameron for the possible mechanisms of jet initiation during the "Whole-Sun workshop in Blaise Pascal Institute". W. Cao acknowledges support from US NSF AST-2108235 and AGS-1821294 grants. BBSO operation is supported by NJIT and US NSF AGS-1821294 grant. GST operation is partly supported by the Korea Astronomy and Space Science Institute and the Seoul National University. We thank SDO/HMI, SDO/AIA teams for the free access to the data.  
\end{acknowledgements}


\bibliography{A_Fan_Shaped_Jet.bib}
\bibliographystyle{aa}

\end{document}